\documentclass[prd,eqsecnum,twocolumn,amsfonts,showpacs]{revtex4}

\usepackage{graphicx}

\usepackage{bm}

\newcommand{\beq}{\begin{equation}}
\newcommand{\eeq}{\end{equation}}
\newcommand{\beqs}{\begin{eqnarray}}
\newcommand{\eeqs}{\end{eqnarray}}

\begin{document}

\title{On Properties of the Ising Model for Complex Energy/Temperature and 
Magnetic Field}

\author{Victor Matveev}

\affiliation{
Department of Mathematical Sciences \\
New Jersey Institute of Technology \\
Newark, NJ  17102}

\author{Robert Shrock}

\affiliation{
C.N. Yang Institute for Theoretical Physics \\
State University of New York \\
Stony Brook, NY 11794}

\begin{abstract}

We study some properties of the Ising model in the plane of the complex
(energy/temperature)-dependent variable $u=e^{-4K}$, where $K=J/(k_BT)$, for
nonzero external magnetic field, $H$.  Exact results are given for the phase
diagram in the $u$ plane for the model in one dimension and on infinite-length
quasi-one-dimensional strips.  In the case of real $h=H/(k_BT)$, these results
provide new insights into features of our earlier study of this case.  We also
consider complex $h=H/(k_BT)$ and $\mu=e^{-2h}$.  Calculations of complex-$u$
zeros of the partition function on sections of the square lattice are
presented.  For the case of imaginary $h$, i.e., $\mu=e^{i\theta}$, we use
exact results for the quasi-1D strips together with these partition function
zeros for the model in 2D to infer some properties of the resultant phase
diagram in the $u$ plane.  We find that in this case, the phase boundary ${\cal
B}_u$ contains a real line segment extending through part of the physical
ferromagnetic interval $0 \le u \le 1$, with a right-hand endpoint $u_{rhe}$ at
the temperature for which the Yang-Lee edge singularity occurs at $\mu=e^{\pm
i\theta}$.  Conformal field theory arguments are used to relate the
singularities at $u_{rhe}$ and the Yang-Lee edge.

\end{abstract}



\maketitle

\vspace{16mm}

\newpage
\pagestyle{plain}
\pagenumbering{arabic}

\section{Introduction}
\label{intro}

       The Ising model serves as a prototype of a statistical mechanical system
which undergoes a phase transition in the ${\mathbb Z}_2$ universality class
with associated spontaneous symmetry breaking and long-range order.  At
temperature $T$ on a lattice $\Lambda$ in an external field $H$, this model is
defined by the partition function $Z=\sum_{\{\sigma_j \} } e^{-\beta {\cal
H}}$, with Hamiltonian
\beq
{\cal H} = -J \sum_{<jj'>} \sigma_j \sigma_{j'} - H \sum_j \sigma_j \ , 
\label{ham}
\eeq
where $\sigma_j = \pm 1$ are the classical spin variables on each site $j \in
\Lambda$, $\beta = (k_BT)^{-1}$, $J$ is the spin-spin exchange constant, and
$\langle j j' \rangle$ denote nearest-neighbour sites.  We use the notation
$K = \beta J$, $h = \beta H$,
\beq
u = e^{-4K} \ , \quad \mu=e^{-2h} \ . 
\label{zuv}
\eeq
The free energy is $F = -k_BT f$, where the reduced free energy is $f=\lim_{n
\to \infty} n^{-1} \ln Z$, with $n$ being the number of lattice sites in
$\Lambda$.  Physical realizations of the Ising model include uniaxial magnetic
materials, structural transitions in binary alloys such as $\beta$ brass, and
the lattice-gas model of liquid-gas phase transitions.  The two-dimensional
version of the model was important partly because it was amenable to exact
solution in zero external magnetic field and the critical point was
characterized by exponents that differed from mean-field theory
(Landau-Ginzburg) values \cite{ons}-\cite{mwbook}.  Although we shall phrase
our discussion in the language of the Ising model as a magnetic system, the
results have analogues in the application as a lattice-gas model of a
liquid-gas phase transition, in which $e^{-2h}$ corresponds to the fugacity.

Just as one gains a deeper understanding of functions of a real variable in
mathematics by studying their generalizations to functions of a complex
variable, so also it has been useful to study the generalization of $h$ and $K$
from real to complex values, as was pioneered by Yang and Lee \cite{yl} and
Fisher \cite{fisher65}, respectively.  In this context, one finds that the
values of $u$ where the model has a paramagnetic-to-ferromagnetic (PM-FM) phase
transition and a paramagnetic-to-antiferromagnetic (PM-AFM) phase transition
occur where certain curves in the complex-$u$ plane cross the positive real-$u$
axis. These curves define boundaries ${\cal B}_u$ of complex-$u$ extensions of
the physical phases of the model and arise via the accumulation of zeros of the
partition function in the thermodynamic limit.  One of the aspects of
complex-$u$ singularities studied in early work was their effect on the
convergence of low-temperature series expansions \cite{dg}.  Although the
two-dimensional Ising model has never been solved exactly in an arbitrary
nonzero external magnetic field $H$, the free energy and magnetization have
been calculated for the particular imaginary values $h = i(2\ell+1)\pi/2$ with
$\ell \in {\mathbb Z}$, which map to the single value $\mu=-1$
\cite{yl,mw67,linwu}. In previous work we presented exact determinations of the
boundaries ${\cal B}_u$ for this $\mu=-1$ case on the square, triangular, and
honeycomb lattices, as well as certain heteropolygonal lattices
\cite{ih,only,yy}.  We investigated the complex-$u$ phase diagram of the Ising
model on the square lattice for physical external magnetic field in
\cite{only}, using calculations of partition function zeros and analyses of
low-temperature, high-field (small-$|u|$, small-$|\mu|$) series to study
certain singularities at endpoints of lines or curves of zeros. In that work we
considered real $h$ and the complex set $h=h_r + \ell i \pi/2$ (where $\ell \in
{\mathbb Z}$) that yield real $\mu$.  (Our notation $\mu=e^{-2h}$ follows that
in the original papers on series expansions \cite{tlow,be} that we used in
\cite{only} and should not be confused with the chemical potential in a
liquid-gas context.)

In this paper we continue the study of the complex-$u$ phase diagram of the
Ising model for nonzero $h$.  We present exact results for lattice strips,
including their infinite-length limits, and calculations of partition function
zeros on finite sections of the square lattice.  For the special case of real
$h$ and the subset of complex $h$ that yield real $\mu$, our exact results for
these strips provide new insight into properties that we found for the 2D Ising
model in \cite{only}.  Complex-$u$ zeros of the 2D Ising model partition
function with nonzero field have been studied further in subsequent works
\cite{kim,newp}. Among general complex values of $h$, we pay particular
attention to the case where $h$ is pure imaginary.  This is of interest partly
because of an important property of the Ising model that was proved by Yang and
Lee \cite{yl}, namely that for the ferromagnetic case ($J>0$), the zeros of the
partition function in the $\mu$ plane lie on the unit circle $|\mu|=1$, i.e.,
correspond to imaginary $h$.  In the limit where the number of sites $n \to
\infty$, these zeros merge to form the locus ${\cal B}_\mu$ comprised of a
connected circular arc $\mu=e^{i\theta}$, where $i\theta = -2h$, passing
through $\mu=-1$ (i.e., $\theta=\pi$) and extending over on the right to a
complex-conjugate pair of endpoints at $e^{\pm i\theta_e}$.  This result
applies for the Ising model in any dimension; indeed, it does not require
$\Lambda$ to be a regular lattice. One interesting question that we address is
the following: what is the phase boundary ${\cal B}_u$ in the $u$ plane for
$\mu = e^{i\theta}$ when $\theta$ is not equal to one of the two exactly solved
cases, i.e., $\theta \ne 0$ mod $\pi$.  We answer this question with exact
results for quasi-1D strips and study it with partition function zeros for 2D.
We find that, in general, ${\cal B}_u$ contains a real line segment extending
through part of the physical ferromagnetic interval $0 \le u \le 1$, with a
right-hand endpoint $u_{rhe}$ at the temperature for which the Yang-Lee edge
singularity occurs at $\mu=e^{\pm i\theta}$. We use conformal field theory
arguments to relate endpoint singularities in the $u$ plane for real and
imaginary $h$ to the Yang-Lee endpoint (edge) singularity.

\section{Relevant Symmetries}

   We record here some basic symmetries which will be used in our work.  On a
lattice with even (odd) coordination number, the Ising model partition function
$Z$ is a Laurent polynomial, with both positive and negative powers, in $u$ (in
$\sqrt{u}$).  $Z$ is also a Laurent polynomial in $\mu$ and hence, without loss
of generality, we consider only the range
\beq
 -\frac{i\pi}{2} < {\rm Im}(h) \le \frac{i\pi}{2} \ . 
\label{imhrange}
\eeq
Furthermore, $Z$ is invariant under the simultaneous transformations $h \to
-h$, $\sigma_j \to -\sigma_j \forall j \in \Lambda$.  The sign flip $h \to -h$
is equivalent to the inversion map
\beq
h \to -h \ \leftrightarrow \ \mu \to \frac{1}{\mu} \ . 
\label{uinversion} 
\eeq
Hence, in considering nonzero real $h$, one may, with no loss of generality,
restrict to $h \ge 0$.  More generally, in considering complex $h$, one may,
with no loss of generality, restrict to the unit disk in the $\mu$ plane,
$|\mu| \le 1$.  It is of particular interest to consider two routes in the
complex $\mu$ plane that connect the two values of $\mu$ where the 2D Ising
model has been exactly solved, viz., $\mu=1$ ($h=0$) and $\mu=-1$ ($h=i\pi/2$).
The first such route is the one that we used in \cite{only}, viz., the real
segment $-1 \le \mu \le 1$.  A second route proceeds along the unit circle
$\mu=e^{i\theta}$.  In view of the above symmetries, it will suffice to
consider this route as $\theta$ increases from 0 to $\pi$. If $\mu \in {\mathbb
R}$, then the set of zeros of $Z$ in the $u$ plane is invariant under $u \to
u^*$ and hence the asymptotic locus ${\cal B}_u$ is invariant under $u \to
u^*$.

The invariance of the set of complex-temperature zeros in $u$ under the complex
conjugation $u \to u^*$ holds not just for real $\mu$ but more generally for
$\mu$ on the unit circle $|\mu|=1$.  This is proved as follows.  For any
lattice $\Lambda$, $Z(\Lambda;u,\mu)=Z(\Lambda;u,1/\mu)$.  Now if and only if
$\mu=e^{i\theta}$ (with real $\theta$), then $\mu^{-1}=\mu^*$.  Hence, for this
case of $\mu=e^{i\theta}$, 
\beq
\mu=e^{i\theta} \ \Rightarrow \ Z(\Lambda;u,\mu)=
                                Z(\Lambda;u,\mu^*) \ . 
\label{inversionconjugation}
\eeq
Now since $Z(\Lambda,u,\mu^*)=[Z(\Lambda,u^*,\mu)]^*$, it follows that 
if $\mu=e^{i\theta}$ (with $\theta \in {\mathbb R}$), then 
\beq
Z(\Lambda;u,e^{i\theta})=0 \ \Leftrightarrow \ 
Z(\Lambda;u^*,e^{i\theta})=0 \ , 
\label{zerosym_mucircle}
\eeq
so that the set of zeros of the partition function in the $u$ plane is
invariant under complex conjugation for this case.

\section{Properties of 1D Solution} 

\subsection{General} 

Because of its simplicity and exact solvability, the 1D Ising model provides
quite useful insights into properties for complex temperature and field.  As is
well known, the Perron-Frobenius theorem guarantees that the free energy and
thermodynamic quantities of a spin model with short-ranged interactions are
analytic functions for any finite temperature on infinite-length strips of
bounded width.  The physical thermodynamic properties of the Ising model on
quasi-one-dimensional strips are thus qualitatively different from those on
lattices of dimensionality $d \ge 2$.  However, in addition to the physical
critical point at zero temperature, switching of dominant eigenvalues of the
transfer matrix and associated non-analyticity in these quantities can occur
for complex $K$ and/or $h$.  Indeed, the properties of the model, on quasi-1D
strips, for complex $u$ and $\mu$ exhibit some interesting similarities to
those on higher-dimensional lattices.  For example, the phase boundary for the
zero-field Ising model on the square, triangular, and honeycomb lattices
exhibits a multiple point at $u=-1$.  (Here, the term ``multiple point'' is
used in the technical sense of algebraic geometry and is defined as a point
where two or more branches of the curves comprising this boundary cross each
other.)  This feature of the model on the 2D lattices is also found to occur
for quasi-1D strips such as the $L_y=2$ strips of the square \cite{a},
triangular \cite{ta}, and honeycomb \cite{hca} lattices.  Furthermore, the fact
that the circle theorem of \cite{yl} applies in any dimension means that there
is particular interest in using quasi-1D strips to obtain exact results on the
singular locus ${\cal B}_u$ corresponding to a point on the unit
circle in the $\mu$ plane.

\subsection{Calculation of ${\cal B}_u$ and Analysis of
  Thermodynamic Quantities}

We begin our analysis of quasi-one-dimensional lattice strips with the 1D line
with periodic boundary conditions, i.e., the circuit graph, $C_n$.  The 
well-known transfer matrix is
\beq
T_{1D} = \left(\begin{array}{cc}
e^{K+h} & e^{-K} \cr
e^{-K}  & e^{K-h} 
\end{array} \right)
\label{Tcircuit}
\eeq
The eigenvalues of this transfer matrix are 
\beq 
\lambda_{1D,j} = e^K \bigg [ \cosh(h) \pm \Big ( \sinh^2(h) + e^{-4K} 
\Big )^{1/2} \ \bigg ] \ , 
\label{lamcn}
\eeq
where the $+$ and $-$ signs apply for $j=1$ and $j=2$.  The
eigenvalues have branch-point singularities at $u=u_e$, where 
\beq
u_e = -\sinh^2(h) = -\frac{(\mu+\mu^{-1}-2)}{4} \ . 
\label{using}
\eeq
(the subscript $e$ denotes ``endpoint'').  Note that ${\rm
det}(T_{1D})=2\sinh(2K)$, independent of $h$. The partition function
is $Z(C_n) = {\rm Tr}[(T_{1D})^n] = \sum_{j=1}^2 (\lambda_{1D,j})^n$.  We
restrict to even $n$ to avoid frustration in the antiferromagnetic case.  The
reduced free energy is $f=\ln(\lambda_{1D,max.})$, where $\lambda_{1D,dom.}$
denotes the maximal eigenvalue.  Equivalently,
\beq
f = \frac{1}{2} \int_0^{2\pi} \, \frac{d\phi}{2\pi} \, \ln \Big [ 
(\lambda_{1D,1})^2 + 
(\lambda_{1D,2})^2 - 2 \lambda_{1D,1}\lambda_{1D,2}\cos(\phi) \Big ] \ , 
\label{f1d}
\eeq
i.e., explicitly,  
\beq
f = K + \frac{\ln 2}{2} + \frac{1}{2}\int_0^{\pi} \, \frac{d\phi}{\pi} \, 
\ln \Big [\cosh(2h)+u+(u-1)\cos(\phi) \Big ] \ . 
\label{f1dexplicit}
\eeq
The complex-$u$ phase boundary ${\cal B}_u$ is the locus of solutions
in $u$ to the condition that the argument of the logarithm in
eq. (\ref{f1dexplicit}) vanishes, i.e.,
\beq
{\cal B}_u: \quad \cosh(2h) + u + (u-1)\cos(\phi)=0 \ . 
\label{B1d}
\eeq
In accordance with the general discussion of symmetries given above, ${\cal
B}_u$ is symmetric under $h \to -h$, i.e., $\mu \to 1/\mu$.  If and only if
$\mu=\pm 1$, it is also symmetric under $K \to -K$, i.e., $u \to 1/u$.  The
condition (\ref{B1d}) is condition that there is degeneracy in magnitude among
the dominant (and here, the only) eigenvalues of the transfer matrix,
$|\lambda_{1D,1}|=|\lambda_{1D,2}|$.  For real $h$, this is equivalent to the
condition that the argument of the square root in eq. (\ref{lamcn}) is
negative. The locus ${\cal B}_u$ is thus a semi-infinite line segment on the 
negative real axis, 
\beq
{\cal B}_u: \quad u < u_e \quad {\rm for \ real } \ h  \ . 
\label{ucircuit}
\eeq
The right-hand endpoint of this line segment, $u_{rhe} = u_e$, occurs at $u=0$
if and only if $\mu=1$, i.e., $h=0$.  As $|h|$ increases, $u_e$ 
moves to the left along the negative real axis.  For $h=0$, the phase
boundary ${\cal B}_u$ is noncompact in both the $u$ and $1/u$ planes (one
implying the other by the invariance of ${\cal B}_u$ under the inversion map $u
\to 1/u$), but for $h \ne 0$, it is noncompact in the $u$ plane but compact in
the $1/u$ plane.

For complex $h=h_r + i\pi/2$, or equivalently, $-1 < \mu < 0$, the term
$\cosh(h)$ in the eigenvalues (\ref{lamcn}) is imaginary, so the condition that
these eigenvalues be equal in magnitude is the condition that the square root
should be real. Hence,
\beqs
{\cal B}_u: & & \quad u \ge u_e = 
\frac{2+|\mu|+|\mu|^{-1}}{4} \ \ {\rm for} \ \ \mu < 0 \ . \cr\cr
& & 
\label{usegment}
\eeqs
This is a semi-infinite line segment on the positive real axis in the $u$ plane
with left-hand endpoint $u_{\ell he} = u_e$.  For this case of negative real
$\mu$, $u_{\ell he} \ge 1$, and $u_{\ell h e} \to 1^+$ as $\mu+1 \to 0^+$.  As
$|h| \to \infty$, $u_{\ell h e} \to \infty$. Again, in the $1/u$ plane, this is
a finite line segment from 0 to the inverse of the right-hand side of
eq. (\ref{usegment}).

We next determine ${\cal B}_u$ for $\mu=e^{i\theta}$ on the unit circle in the
$\mu$ plane.  In this case, with $u_e=-\sinh^2(h) = \sin^2(\theta/2)$, one has
\beq
{\cal B}_u: \quad u \le \sin^2(\theta/2)  \quad {\rm for} \ \
\mu=e^{i\theta}
\label{usegmentmucircle}
\eeq
This is a semi-infinite line segment whose right-hand endpoint $u_{rhs} = u_e$
occurs in the physical ferromagnetic interval $0 \le u_{rhe} \le 1$, increasing
from $u=0$ at $\theta=0$ to $u=1$ as $\theta$ approaches $\pi$ from below.  For
all values of $\mu$ on the unit circle except for the points $\mu=\pm 1$, the
locus ${\cal B}_u$ is not invariant under $u \to 1/u$.

Finally, for $\mu=-1$, the eigenvalues are equal in magnitude and opposite in
sign so that, with $n=2\ell$ even, $Z =2z^\ell (1-u)^\ell$. (If $n$ were odd,
then $Z$ would vanish).  Since we keep $n$ even, $Z$ vanishes only at the 
point $u=1$, and ${\cal B}_u$ degenerates from a one-dimensional locus to the
zero-dimensional locus at $u=1$.

\subsection{Singularities at $u_e$}

The physical singularities of the zero-field 1D Ising model at $T=0$ are well
known; taking $J > 0$ without loss of generality, the model exhibits
exponential divergences $\chi = \beta e^{2K}$ and $\xi \sim (1/2)e^{2K}$ in the
susceptibility and correlation length, together with a jump discontinuity in
the spontaneous magnetization.  Here we focus on the singularities in
thermodynamic quantities at $u_e$ for nonzero $h$.  The internal energy per
site is
\beq
U = -J \Bigg [ 1 -\frac{2e^{-4K}}{\Big (\cosh(h)+\sqrt{\sinh^2(h)+u} \ \Big )
\sqrt{\sinh^2(h)+u} } \Bigg ]
\label{u1d}
\eeq
As is evident from this or from an explicit calculation of the specific heat
$C_H=(\partial U/\partial T)_H$, for nonzero $h$, $h \ne i(2\ell+1)\pi/2$ with
$\ell \in {\mathbb Z}$, $C_H$ diverges at $u=u_e$ with exponent $\alpha_e' =
3/2$.  (We use primes on $\alpha$ and $\gamma$ for phases with either
explicitly or spontaneously broken ${\mathbb Z}_2$ symmetry.)  Applying the
standard scaling relation $2-\alpha'=d/y_t$ (where $y_t$ is the thermal
exponent) at $u=u_e$ yields $y_t=2$ at this singularity.  For $h =
i(2\ell+1)\pi/2$, i.e., $\mu=-1$, eq. (\ref{u1d}) reduces to
\beq
U(\mu=-1) = -\frac{J}{\tanh(2K)}
\label{u1dmum1}
\eeq
so that
\beq
C_H(\mu=-1) = -\frac{2k_BK^2}{\sinh^2(2K)} \ . 
\label{cmum1}
\eeq
Hence, if $\mu=-1$, whence $u_e=1$, i.e., $K_e=0$, the specific heat is finite
at $u_e$, and $\alpha_e'=0$.  This is the same value that we found for the 2D
Ising model at $\mu=-1$, $u_s=1$ in \cite{ih}, as discussed further below.

The per-site magnetization is
\beq
M = \frac{\sinh(h)}{\sqrt{\sinh^2(h)+u}} \ . 
\label{mag1d}
\eeq
For $h \ne 0$ this diverges at $u=u_e$ with exponent $\beta_e=-1/2$.  The
susceptibility per site, $\chi = \partial M/\partial H$, is 
\beq
\chi = \frac{\beta u \cosh(h)}{(\sinh^2(h) + u)^{3/2}} \ . 
\label{chi1d}
\eeq
For $h \ne 0$ and $h \ne i(2\ell+1)\pi/2$, this diverges at $u=u_e$ with
exponent $\gamma_e' = 3/2$. Applying the scaling relation $\beta+\gamma'=
y_h/y_t$ (where $y_h$ is the magnetic exponent) at $u_e$ and substituting
$y_t=2$ then yields $y_h=2$ at this singularity, so that $y_t=y_h$ at $u_e$.
For $h=i(2\ell+1)\pi/2$, the $\cosh(h)$ factor causes $\chi$ to vanish
identically, so that no exponent $\gamma_e'$ is defined.  Thus, this exactly
solved model shows that, just as the value $h=0$ is obviously special since it
preserves the ${\mathbb Z}_2$ symmetry, so also the values $h=i(2\ell+1)\pi/2$
are special, leading to different values of singular exponents at $u_e$ than
the values at generic nonzero values of $h$.

We denote the density of zeros on ${\cal B}_u$ as $g(u)$.  As $u$ approaches a
singular point $u_s$ on ${\cal B}_u$, the density of zeros is related to the
critical exponent for the specific heat \cite{fisher65,abe}
\beq
g(u) \sim |u-u_s|^{1-\alpha_s'} \quad {\rm as} \ |u-u_s| \to 0 \ . 
\label{galpha}
\eeq
(This exponent would be denoted $1-\alpha$ if it applies at the critical point,
as approached from the physical high-temperature phase.)  The singular point
$u_s$ may be the critical point, $u_c$, as in the case $h=0$, or the arc
endpoints $u_e$ and $u_e^*$ studied in \cite{only} in the presence of a real
nonzero field, and, in our present discussion we are interested in the endpoint
of the line segment at $u_e$.

Let us first consider the boundary ${\cal B}_u$ for physical $h$.  This locus
is the solution to eq. (\ref{B1d}) and the density of zeros is proportional to
$d\phi/du$.  To begin, we consider real $h$.  It is convenient to introduce a
positive variable $u' = -u$.  If one normalizes the density according to
\beq
\int_{u'_e}^{\infty} \, du' \, g(u') = 1 \ ,
\label{guintegral}
\eeq
then
\beq
g(u') = \frac{2\cosh(h)}{\pi (1+u')\sqrt{u'-\sinh^2(h)}} \ . 
\label{guprime}
\eeq
In the neighborhood of a point where the free energy is singular, one can
write, as was done in \cite{only},
\beq
g(u') \sim \Big |1-\frac{u'}{u_e'} \Big |^{1-\alpha_e'} \ , 
\label{gsing}
\eeq
 From the discussion above, one already knows for $h \ne 0$, $\alpha_e'=3/2$ 
at $u_e$ except if $h=(2\ell+1)i\pi/2$, where $\alpha_e'=0$ and 
$g(u)$ vanishes identically, reflecting the 
above-mentioned fact that ${\cal B}_u$ degenerates to a point at $u=u_e=1$. 
These findings are in agreement with the present analysis of the density of
zeros; for $h \ne (2\ell+1)i\pi/2$, expanding eq. (\ref{guprime}) 
as $u'-u'_e \to 0^+$, we have $g(u') \to (2/\pi)/\sqrt{u'-u'_e}$, so 
$1-\alpha_e'=-1/2$, i.e., $\alpha_e'=3/2$.

We next show the close relation between this singular behavior of the density
of zeros on ${\cal B}_u$ as one approaches the endpoint $u_e$ with the singular
behavior of the zeros on ${\cal B}_\mu$ as one approaches the endpoint $\mu_e$
of that locus.  We focus on the case $J > 0$ and $h$ imaginary, for which
${\cal B}_\mu$ is an arc of the unit circle $\mu=e^{i\theta}$ extending
clockwise from $\theta=\pi$ to $\theta=\theta_e$ and counterclockwise from
$\theta=\pi$ to $\theta=-\theta_e$. The density of zeros on ${\cal B}_\mu$,
denoted $g(\theta)$, has the singular behavior at the endpoint
$e^{i\theta_e}$ given by
\beq
g(\theta) \sim (\theta-\theta_e)^{\sigma}  \quad {\rm as} \ \ 
\theta-\theta_e \to 0^+ \ . 
\label{yl_edge}
\eeq
With $i\theta=-2h$, eq. (\ref{B1d}) becomes
\beq
\cos \theta + u + (u-1)\cos \phi = 0 \ . 
\label{B1dyl}
\eeq
Letting $\phi$ range from 0 to $2\pi$, one sees that the endpoints occur at
\beq
\theta_e = \arccos(1-2u) = 2 \arcsin(\sqrt{u} \ ) \ , 
\label{thetae}
\eeq
i.e., in terms of $\mu$, 
\beq
\mu_e, \mu_e^* = 1-2u \pm 2i \sqrt{u(1-u)} \ . 
\label{mue}
\eeq
The density of zeros, i.e., the number of zeros $N_z$ between $\theta$ and
$\theta+d\theta$, is given by differentiating with respect to $\phi$ and noting
that the totality of these zeros corresponds to the range $0 \le \phi \le
2\pi$: $g(\theta) = dN_z/d\theta = (2\pi)^{-1}d\phi/d\theta$. 
The density is \cite{yl}
\beq
g(\theta) = \frac{\sin(\theta/2)}{2\pi \sqrt{\sin^2(\theta/2)-u}}
\label{gyl}
\eeq
for $\sin^2(\theta/2) > u$ and $g(\theta)=0$ for $\sin^2(\theta/2) < u$.

This density diverges as $\theta - \theta_e \to 0^+$, with the Yang-Lee edge
exponent $\sigma=-1/2$ \cite{yl,fisher78}.  Given the scaling relations 
\beq
\sigma = \frac{d-2+\eta}{d+2-\eta} = \frac{d}{y_h}-1 \ , 
\label{sigma}
\eeq
the result $\sigma_e=-1/2$ is equivalent to $y_h=2$ at $u_e$.  As was noted in
\cite{yl}, for the antiferromagnet ($J < 0$), the zeros in the $\mu$ plane form
a line segment on the negative real $\mu$ axis.  The singularities in the
density $g(\mu)$ at the endpoints of this line segment are again square root
singularities. Thus, for this exactly solved 1D model,
\beq
1-\alpha_e' = \sigma = -\frac{1}{2}  \quad (1D) \ . 
\label{1as}
\eeq
That is, the exponent $1-\alpha_e'$ describing the singular behavior in the
density of partition function zeros in the locus ${\cal B}_u$ in the $u$ plane
as one approaches the endpoint $u_e$ of this locus is the same as the exponent
$\sigma=-1/2$ describing the singular behavior in the density of zeros in the
locus ${\cal B}_\mu$ as one approaches the endpoints of this locus in the $\mu$
plane.  This shows, as we have emphasized in our earlier work
\cite{ih,only,yy}, the value of analyzing the singular locus ${\cal B}$,
including its slice ${\cal B}_u$ in the $u$ plane for fixed $\mu$ and its slice
${\cal B}_\mu$ in the $\mu$ plane for fixed $u$, in a unified manner.  Indeed,
the value of such a unified approach to this singular locus was recognized in
general in early works such as \cite{blomberg1,blomberg2}.  We note also that
for the 1D Ising (and Potts) models, there is a duality relation connecting
temperature and field variables \cite{suz67,glumac}.  The intertwined relation
of the two relevant variables $K$ and $h$ for a (bi)critical point is at the
heart of the analysis of the scaling limit $(T-T_c) \to 0$, $H \to 0$ in terms
of the scaling variable $h/(T-T_c)^{y_h/y_t}$, where $y_t$ and $y_h$ denote the
thermal and magnetic exponents.  A difference is that in the present analysis,
the singular point(s) $u_s$ is (are) not, in general, the physical critical
point.

For the case of complex $\mu$ with $|\mu| \ne 1$, our analysis of the
singularity at $u_e$ goes through as before.  However, in this case, because it
is no longer true that $\mu^{-1}=\mu^*$, the coefficients of the powers of $u$
in the Laurent polynomial comprising $Z$ are not real, so the set of zeros in
the $u$ plane for a given $\mu$ is not invariant under complex conjugation.

\section{Exact Solution for Toroidal Ladder Strip}

\subsection{General Calculation}

Here we consider the ladder strip of the square lattice with doubly periodic
(i.e., toroidal) boundary conditions.  These boundary conditions have the
advantage of minimizing finite-size effects.  They also have the merit that for
any length $L_x$, all of the sites on the lattice have the same coordination
number, equal to the value of 4 for the infinite square lattice.  The periodic
transverse boundary conditions entail a double bond between the sites on the
upper and lower sides of the ladder.  For a given length $L_x$, the strip has
$n=2L_x$ sites.  In the basis
\beq
\Big \{ {+ \choose +}, \quad 
       {+ \choose -}, \quad 
       {- \choose +}, \quad 
       {- \choose -}  \Big \} 
\label{Ly2basis}
\eeq
the transfer matrix for this toroidal ladder ($t\ell$) strip is 
\beq
T_{t\ell} = \left(\begin{array}{cccc}
e^{4K+2h} & e^h    & e^h     &  1         \cr
e^h       & 1      & e^{-4K}  &  e^{-h}    \cr
e^h       & e^{-4K} & 1       &  e^{-h}    \cr
1         & e^{-h} & e^{-h}  &  e^{4K-2h}  
\end{array} \right)
\label{Ttor2}
\eeq
The determinant is 
\beq
{\rm det}(T_{t\ell})=  [2 \sinh(2K)]^4  = \frac{(1-u)^4}{u^2} \ , 
\label{detTtor2}
\eeq
evidently independent of $h$. The partition function is 
\beq
Z_{t\ell}={\rm Tr}[(T_{t\ell})^{L_x}] = \sum_{j=1}^4 (\lambda_{t\ell,j})^{n/2} 
\label{ztor2}
\eeq
where 
\beq
\lambda_{t\ell,1} = 1-u
\label{lam2tbc1}
\eeq
and the three other $\lambda_{t\ell,j}$'s are roots of the cubic equation 
\beq
\lambda^3 + a_2 \lambda^2 + a_1 \lambda + a_0 = 0 
\label{cubeq}
\eeq
where
\beq
a_2 = -\Big ( 1 + u + u^{-1}(\mu + \mu^{-1}) \Big )
\label{a2}
\eeq
\beq
a_1 = \frac{(1-u)\Big [ 1 + u(\mu + \mu^{-1} + 1) \Big ]}{u^2}
\label{a1}
\eeq
and
\beq
a_0 = \frac{(u-1)^3}{u^2} \ . 
\label{a0} 
\eeq
The reduced free energy is given by $f=(1/2)\ln(\lambda_{t\ell,max})$.  Because
of the cumbersome form of the solutions to the cubic equation, we do not
display the explicit results for thermodynamic quantities such as the specific
heat, magnetization, and susceptibility.  Our primary purpose in the analysis
of this toroidal strip is to determine the boundary ${\cal B}_u$ for a given
$\mu$ and to glean some insights from exact results on this boundary for the
case of the model in 2D.

\subsection{Properties at Some Special Points}

In general, a point $(u,\mu)$ is contained in the singular locus ${\cal B}_u$
if there is a switching of dominant eigenvalues of the transfer matrix.  We can
thus immediately derive some results on this locus by considering some special
cases.  

For $u=1$, the eigenvalues are $\lambda_{t\ell,j}= 0$ for $j=1,2,3$ and
$\lambda_{t\ell,4}=(1+\mu)^2/\mu$, so the eigenvalues are equal at this point
if and only if $\mu=-1$.  For this value, they all vanish, as does the
partition function.  Hence, the zero of the partition function at this point
has a multiplicity of $n/2$.  For the quasi-1D strips considered here, this
point $(u,\mu)=(1,-1)$ occurs where ${\cal B}_u$ degenerates to a point. In
contrast, for the square lattice, it is contained as part of a one-dimensional
locus ${\cal B}_u$ \cite{ih}.  

For $u=-1$, the eigenvalues are $\lambda_{t\ell,j}=2$, $j=1,2$ and
\beq
\lambda_{t\ell,j} = -\frac{1}{2\mu} \Big [ (1+\mu)^2 \pm 
\sqrt{(\mu-1)^2(1+6\mu+\mu^2)} \ \Big ]
\label{lamtbc2_um1_34}
\eeq
where the $\pm$ sign applies for $j=3,4$.  
For $\mu=\pm 1$, all four of these eigenvalues have magnitudes equal to 2, 
so the points
\beq
(u,\mu)=(-1,\pm 1) \ \in \ {\cal B} \ . 
\label{points1}
\eeq

For real $\mu > 0$, $\mu \ne 1$, $|\lambda_{t\ell,3}|$ is smaller
than 2, decreasing to 0 as $\mu \to 0$ or $\mu \to \infty$, while 
$|\lambda_{t\ell,4}|$ is larger than 2, approaching infinity as 
$\mu \to 0$ or $\mu \to \infty$, so that there are no further switchings of
dominant eigenvalues for these values of $\mu$. 
We next consider the real interval $\mu < 0$.  The polynomial in the square
root in eq. (\ref{lamtbc2_um1_34}) is negative for $-(3+2\sqrt{2}) \le \mu \le
-(3-2\sqrt{2})$ and $|\lambda_{t\ell,j}|=2$ for all four $j=1,2,3,4$ for this
interval.  Hence,
\beq
{\cal B}_u \ \supset \{u=-1\} \ \ {\rm for} \ \ 
-(3+2\sqrt{2}) \le \mu \le -(3-2\sqrt{2})
\label{negativemu_linesegment}
\eeq
for this strip.  Although we give the full range of $\mu$, we recall that,
owing to the $\mu \leftrightarrow 1/\mu$ symmetry, it is only necessary to
consider the interior of the disk $|\mu|=1$ since the behavior of ${\cal B}_u$
determined by $|\mu|$ in the exterior of this disk is completely determined by
the values of $\mu$ in the interior.

\subsection{ $\mu=1$} 

We now proceed with our analysis of the complex-$u$ phase diagram for the
infinite-length limit of this toroidal ladder strip for specific values and
ranges of $\mu$. For the zero-field case $\mu=1$, the three eigenvalues in
addition to $\lambda_{t\ell,1}$, are
\beq
\lambda_{t\ell,2}=u^{-1}-1 
\label{lamb2tbc2}
\eeq
and
\beq
\lambda_{t\ell,j} = \frac{1}{2} \Big (u+u^{-1}+2 \pm \sqrt{u^2+u^{-2}+14} 
\ \Big ) 
\label{lam2tbc34}
\eeq
where the $\pm$ sign applies for $j=3,4$, respectively.  For this case,
under the symmetry transformation $K \to -K$, the first two eigenvalues are
permuted according to $\lambda_{t\ell,1} \to -\lambda_{t\ell,2}$,
$\lambda_{t\ell,2} \to -\lambda_{t\ell,1}$, while the last two,
$\lambda_{t\ell,3}$ and $\lambda_{t\ell,4}$, are individually invariant.

\begin{figure}
\begin{center}
\includegraphics[height=6cm]{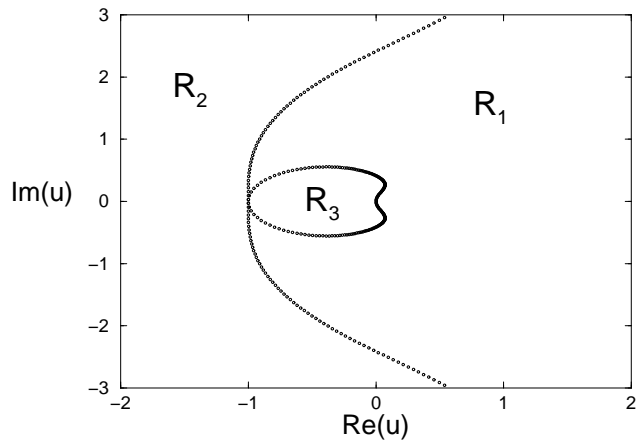}
\end{center}
\caption{\footnotesize{Complex-temperature phase boundary ${\cal B}_u$ and
partition function zeros in the $u$ plane, for the Ising model with $h=0$,
i.e., $\mu=1$, on an $L_y=2$ strip of the square lattice with toroidal boundary
conditions. Zeros are shown for $L_x=200$.}}
\label{tormu1}
\end{figure}

For this $h=0$ case, in the limit $L_x \to \infty$, the boundary ${\cal B}_u$
consists of an inner closed curve shaped like a lima bean passing through the
origin $u=0$ where it has an involution, and through the point $u=-1$.  The
rest of ${\cal B}_u$, which is related to this inner part by the $u \to 1/u$
symmetry, passes through $u=-1$ and extends to $u = \pm i \infty$.  The point
$u=-1$ is an multiple point of osculation type, where the inner and outer
curves on ${\cal B}_u$ coincide with equal (vertical) tangent.  The locus
${\cal B}_u$ thus separates the $u$ plane into three regions, which include the
respective three intervals of the real axis: (i) $R_1$: \ $u \ge 0$, where
$\lambda_{t\ell,3}$ is the dominant eigenvalue; (ii) $R_2$: \ $u < -1$, where
$\lambda_{t\ell,1}$ is dominant; and (iii) $R_3$: \ $-1 \le u \le 0$, where
$\lambda_{t\ell,2}$ is dominant.  Thus, the outer curve is the solution locus
of the equation $|\lambda_{t\ell,1}|=|\lambda_{t\ell,3}|$, while the inner
bean-shaped curve is the solution locus of the equation
$|\lambda_{t\ell,2}|=|\lambda_{t\ell,3}|$. The outer curves cross the imaginary
axis at $u=\pm (\sqrt{2}+1)i$, while the inner curves cross at the inverses of
these points, $u=\mp (\sqrt{2}-1)i$.  In Fig. \ref{tormu1} we show a plot of
complex-temperature zeros calculated for a long finite strip, which clearly
indicate the asymptotic locus ${\cal B}_u$.

\subsection{ $0 \le \mu < 1$} 

\begin{figure}
\begin{center}
\includegraphics[height=6cm]{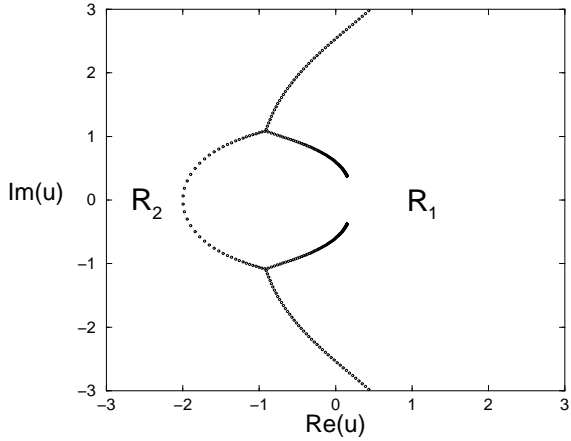}
\end{center}
\caption{\footnotesize{Complex-temperature phase boundary ${\cal B}_u$ and
partition function zeros in the $u$ plane, for the Ising model with $h=(1/2)\ln
2$, i.e., $\mu=1/2$, on a ladder strip with toroidal boundary conditions. 
Zeros are shown for $L_x=200$.}}
\label{tormu0p5}
\end{figure}

We next consider nonzero $h$, recalling that, owing to the symmetries of $Z$
under $h \to -h$, we can, without loss of generality, restrict to the rim and
interior of the unit disk $|\mu| \le 1$ in the $\mu$ plane.  As $h$ increases
from zero through real values, i.e., $\mu$ decreases from 1, the part of the
locus ${\cal B}_u$ that passed through $u=0$ for $h=0$ breaks apart into two
complex-conjugate arcs whose endpoints move away from the real axis.  The
outer curves on ${\cal B}_u$ continue to extend to infinity in the $u$ plane,
passing through the origin $1/u=0$ of the $e^{4K}$ plane.  This is a
consequence of the fact that a nonzero (finite) external magnetic field does
not remove the critical behavior associated with the zero-temperature PM-AFM
critical point of the Ising model on a bipartite quasi-one-dimensional
infinite-length strip.  The locus ${\cal B}_u$ continues to intersect the
negative real axis, at the point
\beq
u = -\frac{1}{\mu} \ . 
\label{uleft}
\eeq
The outer part of the locus ${\cal B}_u$ is comprised of two
complex-conjugate curves that extend to complex infinity, i.e. pass through
$1/u=0$.  The locus ${\cal B}_u$ separates the $u$ plane into two regions: (i)
region $R_1$, which contains the real interval $-\mu^{-1} \le u \le \infty$,
where the root of the cubic with greatest magnitude is dominant, and (ii)
region $R_2$, which contains the real interval $-\infty \le u \le -\mu^{-1}$,
where $\lambda_{t\ell,1}$ is dominant.  The region $R_3$ that was present for
$h=0$ is no longer a separate region, but instead is contained in $R_1$.  As an
illustration of the case of nonzero $h$, we show in Fig. \ref{tormu0p5} a plot
of the phase diagram for $\mu=1/2$, for which ${\cal B}_u$ crosses the
real $u$ axis at $u=-2$.  For this value of $\mu$, the arc endpoints on ${\cal
 B}_u$ are located at $u \simeq 0.149480 \pm 0.376522i$, which are zeros of the
polynomial 
\beqs
& & 64u^8+128u^7+1252u^6+1864u^5+3448u^4-1060u^3 + \cr\cr
& & +937u^2-108u+36 \ , 
\label{rootpolmu0p5} 
\eeqs
which occurs in a square root in the solution of the cubic equation
(\ref{cubeq}).

The behavior of this exactly solved example provides a simple one-dimensional
model of the more complicated behavior on the square-lattice.  For the 2D case
with any nonzero $h$, the part of the singular locus ${\cal B}_u$ that
intersected the real $u$ axis for $h=0$ at the position of the PM-FM critical
point, $u_{PM-FM}=3-2\sqrt{2}$, breaks open, with the two complex-conjugate
endpoints moving away from the real axis, as shown in Fig. 4 of \cite{only}.
This breaking of the boundary and retraction of the arc endpoints away from the
point $u_{PM-FM}$ is in accord with a theorem that for nonzero (physical) $H$,
the free energy is a real analytic function in an interval from $T=0$ beyond
$T_c$ for the PM-FM transition, i.e., in this case, from $u=0$ along the real
$u$ past the point $u=u_c$ \cite{lp}. For our exactly solved quasi-1D strips,
the PM-FM critical point is at $u=0$, which is thus the analogue of
$u_{PM-FM}$.  So the motion of the right-hand endpoint of the semi-infinite
line segment in eq. (\ref{ucircuit}), moving left, away from the point $u=0$,
as $h$ increases in magnitude from zero (through real values), is analogous to
the motion found in \cite{only} of the arc endpoints away from the real axis.
The ladder strip exhibits a behavior (shown in Fig. \ref{tormu0p5}) even closer
to that which we found in the 2D case, namely the breaking of the curve on
${\cal B}_u$ that passes through the former critical point and the retraction
of the complex-conjugate endpoints on ${\cal B}_u$ from the real axis.

\subsection{ $\mu=-1$} 

We can also use our results to consider the complex-field value 
$\mu=-1$ and the interval $-1 \le \mu \le 0$.  We begin with the value
$\mu=-1$. Here the eigenvalues of the transfer matrix take the simple form
$\lambda_{t\ell,1}= 1-u$ as in eq. (\ref{lam2tbc1}) and, for the three others:
\beq
\lambda_{t\ell,2} = 1-u^{-1}
\label{lamm2tbc_m_2}
\eeq
\beq
\lambda_{t\ell,j} = \frac{(u-1)}{2u} \Big [ u+1 \pm \sqrt{1+6u+u^2} \ \Big ] 
\ , 
\label{lamm34}
\eeq
where the $\pm$ signs apply for $j=3,4$, respectively.  All of these
eigenvalues vanish at $u=1$, so that $Z \sim (u-1)^{L_x} \sim (u-1)^{n/2}$
as $u \to 1$, i.e., $Z$ has a zero of multiplicity $n/2$ at $u=1$.

\begin{figure}
  \begin{center}
    \includegraphics[height=6cm]{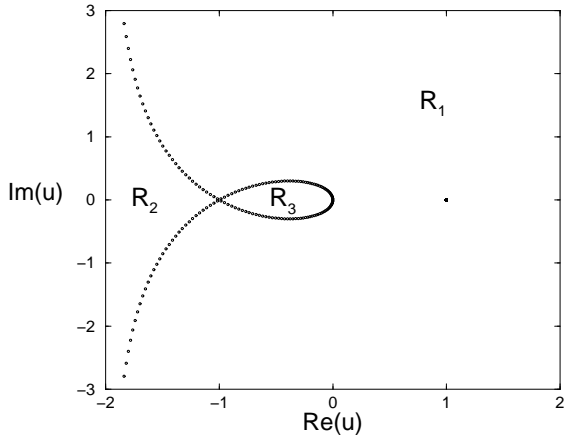}
  \end{center}
\caption{\footnotesize{Complex-temperature phase boundary ${\cal B}_u$ and
partition function zeros in the $u$ plane, for the Ising model with 
$h=\pm i\pi/2$, i.e., $\mu=-1$, on a ladder strip with toroidal boundary
conditions. Zeros are shown for $L_x=150$.}}
\label{tormum1}
\end{figure}

The boundary ${\cal B}_u$ is a curve that passes through the points $u=0$,
$u=-1$, and $1/u=0$, separating the $u$ plane into three regions, as shown in
Fig. \ref{tormum1}: (i) $R_1$, containing the real interval $u \ge 0$, where
$\lambda_{t\ell,3}$ is dominant; (ii) $R_2$, including the real interval $u \le
-1$, where $\lambda_{t\ell,1}$ is dominant; and (iii) $R_3$, the interior of
the loop, including the real interval $-1 \le u \le 0$, where
$\lambda_{t\ell,2}$ is dominant.  There is an isolated point $u=1$ where all
four of the $\lambda_{t\ell,j}$'s, $j=1,...,4$ vanish, and the partition
function itself vanishes.  The invariance of the locus ${\cal B}_u$ under the
inversion map $u \to 1/u$ is evident in Fig. \ref{tormum1}.  The inner loop is
the solution of the equation $|\lambda_{t\ell,2}|=|\lambda_{t\ell,3}|$, while
the outer curve extending to $u = \pm i \infty$ is the solution of the equation
$|\lambda_{t\ell,1}|=|\lambda_{t\ell,3}|$.

\subsection{$-1 < \mu < 0$}

\begin{figure}
  \begin{center}
    \includegraphics[height=6cm]{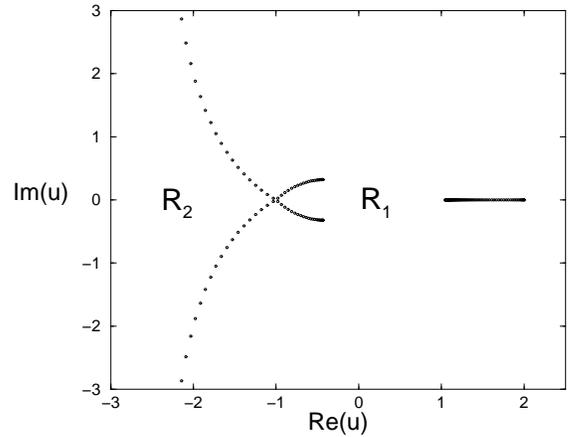}
  \end{center}
\caption{\footnotesize{Complex-temperature phase boundary ${\cal B}_u$ and
partition function zeros in the $u$ plane, for the Ising model with 
$\mu=-1/2$, on a ladder strip with toroidal boundary
conditions. Zeros are shown for $L_x=100$.}}
\label{tormum0p5}
\end{figure}

As $\mu$ increases from $-1$ toward zero through real values, the
above-mentioned loop on the $\mu=-1$ locus ${\cal B}_u$ breaks, with its two
complex-conjugate arcs retracting from $u=0$.  These arcs cross each other at
$u=-1$ with the outer parts continuing to extend upward and downward to
infinity in the $u$ plane, passing through $1/u=0$.  The boundary ${\cal B}_u$
separates the $u$ plane into two regions, $R_1$ to the right, and $R_2$ to the
left, of these semi-infinite arcs.  The single zero with multiplicity $n/2$
that had existed at $u=1$ for $\mu=-1$ is replaced by a finite line segment in
the region $\mu \ge 1$.  As $\mu$ moves to the right from $-1$ toward zero, the
real line segment also moves to the right.  In Fig. \ref{tormum0p5} we show a
plot of zeros for a typical value in this range, $\mu=-1/2$.  For this case,
the arc endpoints in the $L_x \to \infty$ limit occur at approximately $u
\simeq -0.431214 \pm 0.3218815i$ and the real line segment occupies the
interval $1.051945 \le u \le 2$.  These are certain zeros of the polynomial
\beqs
& & (u-2)(64u^7+256u^6-796u^5+ \cr\cr
& & +272u^4+352u^3+4u^2-135u-18) \cr\cr
& & 
\label{srootmum0p5}
\eeqs
that occurs in a square root in the solution of eq. (\ref{cubeq}) for this
case.  The line segment that we find on the real $u$ axis for $-1 < \mu < 0$ is
the analogue of the two line segments on the real $u$ axis that we found for
this range of $\mu$ for the model on the square lattice in \cite{only} (as
shown in Fig. 6 of that reference).

\subsection{ $\mu=e^{i\theta}$ }

Here we analyze the complex-temperature phase diagram for this strip in the
case where $h$ is pure imaginary, i.e., $\mu=e^{i\theta}$.  We show that the
endpoints of the unit-circle arc on ${\cal B}_\mu$, i.e., the Yang-Lee edge
singularities, have a corresponding feature in ${\cal B}_u$, namely an
endpoint of a real line segment that lies in the interval $0 < u < 1$.  For
$\mu=e^{i\theta}$ the eigenvalues of the transfer matrix consist of
$\lambda=1-u$ as in eq. (\ref{lam2tbc1}) and the three roots of the cubic
(\ref{cubeq}).  The coefficients $a_2$ and $a_1$ in this cubic can be expressed
conveniently as
\beq
a_2 = -(1+u+2u^{-1}\cos\theta)
\label{a2mucircle}
\eeq
and
\beq
a_1 = u^{-2}(1-u)\Big [ 1+u(2\cos\theta+1) \Big ] \ . 
\label{a1mucircle}
\eeq
We find that for $\mu$ on the unit circle, the complex-temperature phase
boundary always passes through the points $u=0$, $u=-1$, and $u=\pm i\infty$
(the last corresponding to the single point $1/u=0$ in the plane of the
variable $1/u=e^{4K}$).  We now prove these results. To show that the point
$u=-1$ is on ${\cal B}_u$, we observe that for $u=-1$, the eigenvalue given by
eq. (\ref{lam2tbc1}) has the value $\lambda=2$, and the cubic equation for the
other three eigenvalues factorizes according to
\beq
(\lambda-2)[\lambda^2 + 4\cos^2(\theta/2) \, \lambda + 4]=0 \ , 
\label{lpum1}
\eeq
so that these three other eigenvalues are $\lambda=2$ and 
\beq
\lambda = 2\Big [ -\cos^2(\theta/2) \pm i \sin(\theta/2)
\, \sqrt{1+\cos^2(\theta/2)} \ \Big ] \ . 
\label{rootsum1}
\eeq
All of these have magnitude 2, which proves that the point $u=-1$ is on ${\cal
B}_u$.  Indeed, this calculation shows, further, that four curves on ${\cal
B}_u$ intersect at $u=-1$.  To prove that the point $u=0$ is on ${\cal B}_u$,
we first note that for this value of $u$, the eigenvalue
$\lambda_{t\ell,1}=1-u$ has the value 1.  We multiply eq. (\ref{cubeq}) by
$u^2$ and then take the limit $u \to 0$, obtaining the equation $\lambda-1=0$.
This proves the result since we then have two degenerate dominant eigenvalues.
The same method enables one to conclude that the point $1/u=0$ is on ${\cal
B}_u$.

For any $\theta \ne 0$ mod $\pi$, the locus ${\cal B}_u$ includes a line
segment that occupies the interval $-1 \le u \le 0$ and also occupies part of
the interval [0,1).  We denote the right-hand end of this line segment as
$u_{t\ell,rhe}$. This right-hand endpoint increases monotonically from 0 to 1
as $\theta$ increases from 0 to $\pi$.

\begin{figure}
  \begin{center}
    \includegraphics[height=6cm]{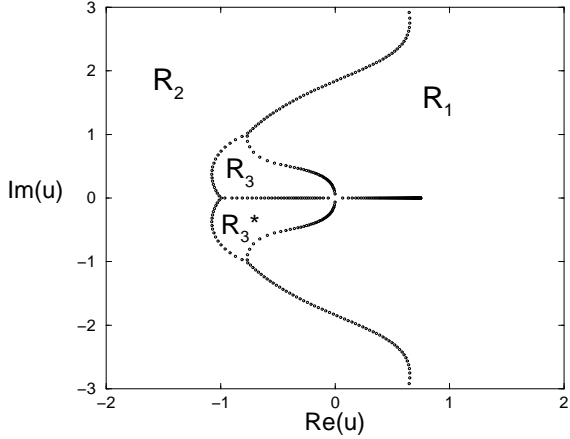}
  \end{center}
\caption{\footnotesize{Phase boundary ${\cal B}_u$ in the $u$ plane for the
Ising model with $\mu=i$, on the ladder strip with toroidal boundary
conditions. Zeros are shown for $L_x=200$.}}
\label{tormui}
\end{figure}

As an illustration of the complex-temperature phase diagram for $\mu$ on the
unit circle, we consider the case $\theta=\pi/2$, i.e., $\mu=i$, for
which the cubic equation (\ref{cubeq}) takes the form
\beqs
& & \lambda^3 - (1+u)\lambda^2 + u^{-2}(1-u^2)\lambda + u^{-2}(u-1)^3 = 0 \ .
\cr\cr
& & 
\label{torcpmui}
\eeqs
In Fig. \ref{tormui} we show the resultant complex-$u$ phase diagram.  The
boundary has a multiple point at $u=0$ where a real line segment intersects a
vertical branch of the curve on ${\cal B}_u$. There are three triple points on
${\cal B}_u$, namely the one at $u=-1$ and a complex-conjugate pair in the
second and third quadrants.  The right-hand of the real line segment occurs at
\beq
u_{t\ell,rhe} \simeq 0.746125 
\label{u_rhe_tor_mui}
\eeq
which is the unique real positive root of the polynomial
\beqs
& & u^8+2u^7-2u^6+26u^5-48u^4+30u^3-2u^2-2u-1 \cr\cr
& & 
\label{upolmui}
\eeqs
that occurs in a square root in the exact solution of the cubic.  ${\cal B}_u$
also includes two complex-conjugate curves that extend upward and downward to
$\pm i \infty$ within the first and fourth quadrants, passing through $1/u=0$.
As is evident from Fig. \ref{tormui}, the boundary ${\cal B}_u$ separates the
$u$ plane into four regions: $R_1$ and $R_2$, extending infinitely far to the
right and left, and the two complex-conjugate enclosed phases separated
by the part of the real line segment $-1 \le u \le 0$.  Qualitatively similar
results hold for other values of $\mu=e^{i\theta}$ with $0 < \theta < \pi$.  In
Table \ref{urhe_theta} we show the values of $u_{rhe,t\ell}$ and the
corresponding values of $k_BT/J$, denoted as $k_BT_{rhe,t\ell}/J$, as functions
of $\theta$.  These are compared with the values for the Ising model on the
infinite line (with periodic boundary conditions), denoted, respectively, as
$u_{rhe,1D}$ and $k_BT_{rhe,1D}/J$.

\begin{table}[htbp]
\caption{\footnotesize{Values of $u_{rhe}=e^{-4K_{rhe}}$ and $k_BT_{rhe}/J =
  K_{rhe}^{-1}$ as a function of $\theta$, with $\mu=e^{i\theta}$, for 1D
  (columns 2,3) and the toroidal lattice ($t\ell$) strip (columns 4,5). We use
  the notation $\bar T \equiv k_B T/J$.}}
\begin{center}
\begin{tabular}{|c|c|c|c|c|}
\hline\hline $\theta$ & $u_{rhe,1D}$  & $u_{rhe,t\ell}$ & $\bar T_{rhe,1D}$ &
$\bar T_{rhe,t\ell}$ \\ \hline\hline
  0      &  $-$                           & $-$     & $-$   & $-$   \\ \hline
$\pi/12$ & $(2-\sqrt{2+\sqrt{3}})/4 \simeq 0.017$ 
                                          & 0.1918  & 0.982 & 2.422 \\ \hline
$\pi/6$  & $(2-\sqrt{3})/4 \simeq 0.0670$ & 0.3315  & 1.480 & 3.622 \\ \hline
$\pi/4$  & $(2-\sqrt{2})/4 \simeq 0.1464$ & 0.45245 & 2.082 & 5.044 \\ \hline
$\pi/3$  & 1/4                            & 0.5612  & 2.885 & 6.924 \\ \hline
$\pi/2$  & 1/2                            & 0.7461  & 5.771 &13.658 \\ \hline
$2\pi/3$ & 3/4                            & 0.8846  &13.904 &32.625 \\ \hline
$3\pi/4$ & $(2+\sqrt{2})/4 \simeq 0.85355$& 0.9346  &25.261 &59.107 \\ \hline
$5\pi/6$ & $(2+\sqrt{3})/4 \simeq 0.9330$ & 0.9707  &57.690 &134.725 \\\hline
$\to \pi$ & $\to 1$ & $\to 1$  &  $\to \infty$ & $\to \infty$ \\ \hline\hline 
\end{tabular}
\end{center}
\label{urhe_theta}
\end{table}

\section{Exact Solution for Cyclic Ladder Strip}

We have also carried out a similar study of ${\cal B}_u$ for the ladder strip
with cyclic (i.e., periodic longitudinal and free transverse) boundary
conditions.  As before, we take the length $L_x$ to be even to maintain the
bipartite property of the infinite square lattice.  Strips of this type (with
$L_x \ge 4$ to avoid degeneration) have the property that all of the sites have
the same coordination number, 3.  They are thus not expected to exhibit
properties that are as similar to those of the square lattice as the toroidal
ladder strip (which has the same coordination number as the infinite square
lattice).  Furthermore, since the transverse boundary conditions are free
rather than periodic, finite-size effects are larger for this lattice than for
the toroidal strip, which has no boundaries.  Thus, we include a discussion of
the Ising model on this cyclic ladder strip mainly for comparative purposes, 
but since the strip shares fewer similarities with the infinite square lattice,
our treatment will be more brief than for the toroidal strip. 

Since the cyclic ladder strips have odd coordination number, the partition
function is a Laurent polynomial in $\mu$ and in $x=e^{-2K}$, rather than
$u=x^2=e^{-4K}$. (Here we switch notation from that used in our previous papers
\cite{chisq,ih,only,yy}, using $x$ rather than $z$ for $e^{-2K}$ in order to
avoid confusion with the fugacity $z=e^{\beta \mu'}$, where $\mu'$ is the
chemical potential.)  In \cite{yy} we proved that for lattices with odd
coordination number, the following equality holds (up to a possible overall
factor that does not affect the zeros): $Z(x,-\mu) = Z(-x,\mu)$.  This theorem
implies that the accumulation set of the zeros satisfies
\beq
{\cal B}(x,-\mu) = {\cal B}(-x,\mu) \ . 
\label{bsymodd}
\eeq
In particular, for real $\mu$, this symmetry, together with the inversion
symmetry $\mu \to 1/\mu$, means that it suffices to consider just the interval
$0 \le \mu \le 1$. In the basis (\ref{Ly2basis}) the transfer matrix is
\beq
T_{c\ell} = \left(\begin{array}{cccc}
e^{3K+2h} & e^h     & e^h      &  e^{-K}    \cr
e^h       & e^K     & e^{-3K}  &  e^{-h}    \cr
e^h       & e^{-3K} & e^K      &  e^{-h}    \cr
e^{-K}    & e^{-h}  & e^{-h}   &  e^{3K-2h}
\end{array} \right)
\label{Tcyc2}
\eeq
where $c\ell$ denotes ``cyclic ladder''.  This transfer matrix has the same
determinant as for the toroidal ladder.  The partition function is given by
\beq
Z_{c\ell} = \sum_{j=1}^4 (\lambda_{c\ell,j})^{L_x}
\label{zcyc2}
\eeq
where $\lambda_{c\ell,j}$, $j=1,..4$ are the eigenvalues of $T_{c\ell}$.  the
reduced free energy is $f=(1/2)\ln(\lambda_{c\ell,max})$.  The eigenvalues are
$\lambda_{c\ell,1}=x^{-1/2}(1-x^2)$ and the three roots of the cubic polynomial
that comprises the rest of the characteristic polynomial of $T_{c\ell}$. The
reduced free energy is $f=(1/2)\ln(\lambda_{c\ell,max})$.

For $h=0$ the phase boundary ${\cal B}_x$ for the $q$-state Potts model and, in
particular, the $q=2$ Ising case, was analyzed in \cite{a}.  This boundary
consists of curves that pass through $x=0$, $x=\pm i$, and $1/x=0$ with each of
these points being a multiple point where two branches of ${\cal B}_x$ cross
each other.  The curves separate the $x$ plane into six regions, two of which
include the real intervals $x \le 0$ and $x \le 0$, and the other four of which
include the intervals on the imaginary axis $-\infty \le {\rm Im}(x) \le -1$,
$-1 \le {\rm Im}(x) \le 0$, $0 \le {\rm Im}(x) \le 1$, and $1 \le {\rm Im}(x)
\le \infty$.  There is also an isolated zero of the partition function at
$x=-1$ with multiplicity scaling like $n$.

Here we focus on the case of nonzero $h$.  For this case the two pairs of
complex-conjugate curves connecting $x=0$ with $x=\pm i$ each break, and the
endpoints move away from the real axis as $\mu$ decreases from 1 to 0; at the
same time, the multiple zero at $x=-1$ is replaced by a line segment.  One of
the reasons for studying this lattice strip is to confirm that the singular
locus ${\cal B}_x$ again has a line segment in the physical ferromagnetic
region, just as we found for the 1D line and the toroidal strip.  We do,
indeed, confirm this, showing the generality of this important result.  A
general property of ${\cal B}_x$ for the Ising model on this cyclic strip is
that for $\mu$ on the unit circle, $\mu=e^{i\theta}$, this boundary ${\cal
B}_x$ passes through $x=0$, $1/x=0$, and $x=\pm i$.  For $\mu \ne \pm 1$, there
are line segments on the real axis.  The value $\mu=i$ is especially simple,
since the invariance of ${\cal B}$ under $\mu \to 1/\mu$ and the symmetry for
lattices of odd coordination number, (\ref{bsymodd}), together imply that for
$\mu=i$, ${\cal B}_x$ is invariant under $x \to -x$ and hence can be depicted
in the $u$ plane.  We find that for this case $\mu=i$, the right-hand endpoint
of the real line segment occurs at $x_{rhe} \simeq 0.82942$, or equivalently,
$u_{rhe} \simeq 0.68794$, which is the unique positive root of the polynomial
\beq
u^6+29u^4-48u^3+27u^2-4u-1 \ , 
\label{cycrootpol}
\eeq
which occurs in a square root in the exact solution to the cubic equation for
the eigenvalues of the transfer matrix.  Since the coordination number of this
cyclic lattice is intermediate between the value, 2, for the periodic 1D line
and the value 4 for the toroidal ladder strip, one expects that the value of
$u_{rhe}=x_{rhe}^2$ at a given value of $\theta$ would also lie between those
for the 1D line and the toroidal strip.  This is verified; we find (see Table
\ref{urhe_theta}) the respective values $u_{rhe}=0.5$, 0.6879, and 0.7461, for
the 1D line, and cyclic toroidal ladder strips.  These values increase
monotonically as the strip width increases and can be seen to approach the
value of $u_{rhe} \simeq 0.8$ that we infer for the thermodynamic limit of the
square lattice from our calculations of partition function zeros, to be
discussed below.

\section{Relations Between Complex-$u$ Phase Diagram for the Ising Model in 1D
  and 2D for Real $\mu$}

In this section we give a unified comparative discussion of how our exact
results for ${\cal B}_u$ on quasi-1D strips relate to exact results for ${\cal
B}_u$ in 2D for $\mu=\pm 1$ and the case of real $\mu$ in the interval $-1 <
\mu < 1$ that we studied earlier in \cite{only}. We first review some relevant
background concerning the phase diagram for the two cases where this diagram is
known exactly for the 2D Ising model, namely $\mu=1$ ($h=0$) and $\mu=-1$
($h=i\pi/2$)

\subsection{$\mu=1$}

The complex-$u$ phase boundary ${\cal B}_u$ for the square-lattice Ising model
is the image in the $u$ plane of the circles \cite{fisher65}
\beq
|x \pm 1|=\sqrt{2} \ , 
\label{fishercircles}
\eeq
namely the lima\c{c}on (Fig. 1c of \cite{chisq}) given by
\begin{eqnarray}
{\rm Re}(u) & = & 1 + 2\sqrt{2} \, \cos(\omega) + 2\cos(2\omega) \cr\cr
{\rm Im}(u) & = & 2\sqrt{2} \, \sin(\omega) + 2\sin(2\omega) 
\label{limacon}
\end{eqnarray}
with $-\pi \le \omega \le \pi$. The outer branch of the lima\c{c}on intersects
the positive real-$u$ axis at $u_{PM-AFM}=3+2\sqrt{2}$ (for $\omega=0$) and
crosses the imaginary-$u$ axis at $u=\pm (2 + \sqrt{3} \ )i$ (for $\omega=\pm
5\pi/12$).  The inner branch of the lima\c{c}on crosses the positive real axis
at $u_{PM-FM}=3-2\sqrt{2}$ (for $\omega=\pi$) and the imaginary-$u$ axis at
$u=\mp (2 - \sqrt{3} \ )i$ (for $\omega=\pm 11\pi/12$).  The lima\c{c}on has a
multiple point at $u=-1$ (for $\omega=\pm 5\pi/4$) where two branches of ${\cal
B}_u$ cross each other at right angles.  When $u=-1$, there are also branches
of the lima\c{c}on passing through ${\rm Im}(u) = \pm 2\sqrt{2} \, i$ (for
$\omega=\pm \pi/2$).  The boundary ${\cal B}_u$ separates the $u$ plane into
three phases, which are the complex extensions of the physical PM, FM, and AFM
phases.  Since the infinite-length strips are quasi-1D, the Ising model has no
finite-temperature phase transition on these strips, and is critical only at
$T=0$.  Thus, for $h=0$, the boundary ${\cal B}_u$ passes through $u=0$ and
$1/u=0$.  However, just as for the square lattice, for the toroidal strip the
boundary ${\cal B}_u$ separates the $u$ plane into three regions, as is evident
in Fig. \ref{tormu1}.  One can envision a formal operation on the boundary
curve ${\cal B}_u$ for the toroidal ladder strip that transforms it into the
${\cal B}_u$ for the 2D lattice, namely to move the crossing at $u=0$ to
$u_{PM-FM}$, which, owing to the $u \leftrightarrow 1/u$ inversion symmetry,
automatically means that the part of the boundary ${\cal B}_u$ that goes to
infinity in the $u$ plane is pulled back and crosses the real axis at the
inverse of this point, viz., $u_{PM-AFM}$.

The complex-$u$ phase boundaries ${\cal B}_u$ of the Ising model on both the
infinite-length 1D line and on the infinite-length ladder strip with toroidal
or cyclic boundary conditions have the property that they pass through $u=-1$
and, for the toroidal and cyclic ladder strips this is again a multiple point
on ${\cal B}_u$, just as it is in 2D. For the toroidal strip, the point $u=-1$
is an osculation point, where two branches on ${\cal B}_u$ intersect with the
same tangent, whereas for the square lattice the branches cross at right
angles.  Other similarities include the fact that, e.g., for the toroidal
strip, ${\cal B}_u$ crosses the imaginary $u$ axis at two pairs of complex
conjugate points that are inverses of each other, namely $u=\pm(\sqrt{2}+1)i$
and $u=\pm(\sqrt{2}-1)i$.  These points are in 1--1 correspondence with the
points $u=\pm (2 \pm \sqrt{3} \ )i$ where ${\cal B}_u$ crosses the
imaginary-$u$ axis for the square lattice.

\subsection{$\mu=-1$} 

The phase boundary for the Ising model with $\mu=-1$ on the square lattice was
determined in \cite{ih} and consists of the union of the unit circle and a
line segment on the negative real axis:
\beqs
& & {\cal B}_u(\mu=-1): \ \{|u|=1\} \ \cr\cr
& \cup & \{-(3+2\sqrt{2}) \le u \le -(3-2\sqrt{2}) \} \ .
\label{bumum1}
\eeqs
It is interesting that the endpoints of this line segment are minus the values
of $u_{PM-FM}$ and $u_{PM-AFM}=1/u_{PM-FM}$ on the square lattice. The point
$u=-1$ is a multiple point on ${\cal B}_u$ where the unit circle $|u|=1$
crosses the real line segment at right angles.  The latter feature is matched
by the locus ${\cal B}_u$ for the toroidal ladder strip, as is evident in
Fig. \ref{tormum1}.

\subsection{$0 \le \mu < 1$}

In \cite{only} it was found that as $h$ is increases from 0, i.e., as $\mu$
decreases from 1 to 0, the inner loop of the lima\c{c}on immediately breaks
open at $u=u_{PM-FM}$, forming a complex-conjugate pair of prong endpoints
$u_e, \ u_e^*$ that retract from the real axis. 
In \cite{only} we used calculations of complex-$u$ partition function zeros
together with analyses of low-temperature, high-field series to determine the
locations of these arc endpoints and the values of the exponents $\alpha_e'$,
$\beta_e'$, and $\gamma_e'$ describing the singular behavior of the specific
heat, magnetization, and susceptibility at these prong endpoints. 

In contrast, the PM-AFM critical point does not disappear.  For the
antiferromagnetic sign of the spin-spin coupling, $J < 0$, as $H$ increases,
the N\'eel temperature $T_N$ decreases, or equivalently, $-K_c = |J|/(k_BT_N)$
increases, and hence also $u_{ZM-AFM}$ increases from its value of
$3+2\sqrt{2}$ at $H=0$ (where the notation $ZM$ follows \cite{only}).  As $H$
increases sufficiently, there is a tricritical point, and when it increases
further to $-c \, J = c \, |J|$, where $c$ denotes the
coordination number, the N\'eel temperature is reduced to zero.  This means
that $\beta_{ZM-AFM} \to \infty$.  Thus, asymptotically as $h \to \infty$,
$-K_{ZM-AFM}/h \to c$, i.e., the right-hand side of the boundary ${\cal
B}_u$ moves outward to infinity like $u \sim \mu^{-1/2}$ as $\mu \to 0$.

With the replacement of the finite-temperature PM-FM critical point $u=u_c$ by
the zero-temperature critical point $u=0$, the boundary ${\cal B}_u$ for the
Ising model on the infinite-length limit of the toroidal ladder strip
reproduces this feature of the model in 2D, viz., immediate breaking of the
loop, as is evident in Fig. \ref{tormu0p5}.  The corresponding boundary ${\cal
B}_z$ for the infinite-length limit of the cyclic strip also immediately breaks
apart from the zero-temperature PM-FM critical point at $u=0$. 

\subsection{$-1 \le \mu \le 0$}

For $\mu$ in the interval $-1 < \mu \le 0$, in the 2D case, the partition
function zeros calculated in \cite{only} exhibited patterns from which one
could infer that in the thermodynamic limit the resultant accumulation locus
${\cal B}_u$ exhibited curves and two line segments, one on the positive, and
one on the negative real $u$ axes. (There were also some zeros that exhibited
sufficient scatter that one could not make a plausible inference about the
asymptotic locus in the thermodynamic limit.)  One may thus ask if we obtain
qualitatively similar behavior with these exact closed-form solutions for the
model on quasi-one-dimensional strips. For both the 1D line and the toroidal
ladder strip with this range of $\mu$, we find a line segment on the positive
real axis (cf. eq. (\ref{usegment} and Fig. \ref{tormum0p5}), in agreement with
this feature that we had obtained in 2D.  As could be expected, the quasi-1D
strips do not reproduce all of the features that we found for 2D.  For example,
neither the 1D line nor the toroidal strip exhibits a real line segment on the
negative real axis, either for the case $\mu=-1$ or the range $-1 < \mu < 0$,
where we did find such a line segment in 2D.

\section{Complex-$u$ Phase Diagram and Zeros of the Partition Function for 
the Square Lattice with $\mu=e^{i\theta}$ }

\subsection{Motivation and Exact Results for $\theta=0$ and $\theta=\pi$}

In this section we present our calculations of complex-$u$ zeros of the
partition function of the square-lattice Ising model for imaginary $h$, i.e.,
$\mu = e^{-2h} = e^{i\theta}$ with $0 < \theta < \pi$.  Owing to the
invariance of the model under the inversion (\ref{uinversion}), it suffices to
consider this half-circle.  This study is a continuation of our earlier
investigation in \cite{ih,only} of the complex-$u$ phase diagram of the
model for real nonzero external magnetic fields (hence $0 \le \mu \le \infty$)
and the subset of complex $h$ of the form $h=h_r \pm i\pi/2$ yielding negative
real $\mu$, and thus covering the interval $-\infty \le \mu \le 0$.

As noted above, in studying the complex-$u$ phase diagram, it is natural to
consider paths in the $\mu$ plane that connect the two values for which this
phase diagram is exactly known, namely $\mu=1$ ($h=0$) and $\mu=-1$
($h=i\pi/2$).  In \cite{only} we considered the path defined by the real
interval $-1 \le \mu \le 1$.  Here we concentrate on the other natural path,
namely an arc along the unit circle $\mu=e^{i\theta}$ with $0 < \theta < \pi$.
For $\mu$ on this unit circle we have mentioned above that the boundary ${\cal
B}_u$ is invariant under complex conjugation.  One motivation for studying the
complex-$u$ zeros of the partition function for $\mu$ on the unit circle is
that the latter locus is precisely where the zeros of the complex-$\mu$ zeros
of the partition function occur for physical temperatures in the case of
ferromagnetic couplings.  Hence, our results in this section constitute an
investigation of the pre-image in the $u$ plane, for the square-lattice Ising
model, of points on the Yang-Lee circle.  Indeed, just as we found with exact
results on infinite-length quasi-1D strips, our calculations of partition
function zeros for the model in 2D will lead us to the inference that in the
thermodynamic limit the locus ${\cal B}_u$ for $\mu=e^{i\theta}$ with $\theta
\ne 0$ mod $\pi$ contains a line segment extending into the physical
ferromagnetic region with a right-hand endpoint $u_{rhe}$ that corresponds
precisely to the temperature for which the points $\mu=e^{\pm i\theta}$ are the
endpoints (Yang-Lee edges) of the arc of the unit circle $|\mu|=1$ comprising
${\cal B}_\mu$. A convenient feature for the study of the complex-$u$ phase
diagram for the square-lattice Ising model with $\mu=e^{i\theta}$ is that the
phase boundary ${\cal B}_u$ remains compact throughout the entire range of
$\theta$.  This is in contrast to the situation for the real path $-1 \le \mu
\le 1$.  In that case, as was discussed in \cite{only}, the phase boundary
separating the phase where the staggered magnetization $M_{st}$ vanishes
identically from the AFM phase where $M_{st}$ is nonzero moves outward to
complex infinity as $\mu \to 0$ and then comes inward again as $\mu$ passes
through 0 and approaches $\mu=-1$.  (It should be noted that although ${\cal
B}_u$ is compact for $\mu=e^{i\theta}$ with $\theta \in {\mathbb R}$ for the
square lattice, this is not the case with the triangular lattice.  On that
lattice, for both of the exactly solved cases $\theta=0$ and $\theta=\pi$, the
locus ${\cal B}_u$ contains the respective semi-infinite line segments $-\infty
\le u \le -1/3$ and $-\infty \le u \le -1/2$ \cite{only}.)

In our previous work \cite{only}, we tested several different types of boundary
conditions including doubly periodic (toroidal, TBC) and helical boundary
conditions (HBC).  The latter are periodic in one direction, say $L_x$, and
helical in the other, say $L_y$.  We found that helical boundary conditions
yielded zeros that showed somewhat less scatter for general $\mu$ and were
closer to the exactly known loci ${\cal B}_u$ for the cases $\mu=\pm 1$ than
the zeros obtained with periodic boundary conditions.  This can be
interpreted as a consequence of the fact that for toroidal boundary conditions,
the global circuits around the lattice have length $L_x$ and $L_y$, while for
helical boundary conditions, while the circuit in the $x$ direction is still of
length $L_x$ the one in the $y$ direction is made much longer, essentially
$L_xL_y$. For the present work we have again made use of helical boundary
conditions and also a set of boundary conditions that have the effect of
yielding zeros that lie exactly on the asymptotic loci ${\cal B}_u$ for the
exactly known cases $\mu=\pm 1$.  These are defined as follows.  We consider
two $L_x \times L_y$ lattices, with the $x$ direction being the longitudinal
(horizontal) and and the $y$ direction the transverse (vertical) one.  We
impose periodic longitudinal boundary conditions and fixed transverse boundary
conditions Specifically, we fix all of the spins on the top row to be $+$ while
those on the bottom row alternate in sign as $(+-+-...)$.  For the second
lattice, we impose spins on the top and bottom rows that are minus those of the
first lattice; that is, all spins on the top row are $-$, while those on the
bottom are $(-+-+...)$.  Together, these yield a partition function that is
invariant under the $h \to -h$ symmetry. We denote these as symmetrized fixed
boundary conditions (SFBC).  In passing, we note that if one used only the
first lattice, the corresponding boundary conditions would correspond to set A
of \cite{bk}.  For $h=0$ this set was shown to yield zeros that lie exactly on
the circles (\ref{fishercircles}).  For our present work, the boundary
conditions of \cite{bk} would not be appropriate, since they violate the $h \to
-h$ symmetry and hence also the $\mu \to 1/\mu$ symmetry of the infinite square
lattice.  In turn, this violation would have the undesirable consequence that
for $\mu=e^{i\theta}$, the set of zeros would not be invariant under complex
conjugation and zeros that should be exactly on the real-$u$ axis would not be.
We have found that the symmetrized fixed boundary conditions yield zeros with
somewhat less scatter than helical boundary conditions, and therefore we
concentrate on the former in presenting our results here.

For the analytic calculation of the partition function, we again use a transfer
matrix method similar to that employed in our earlier paper \cite{only}.  In
that work we performed a number of internal checks to confirm the accuracy of
the numerical calculations of the positions of the zeros of the partition
function. Since for our present study we are performing calculations of
partition functions and zeros for considerably larger lattices than we used in
\cite{only}, we have paid special attention to guaranteeing the accuracy of the
numerical solution for these.  Among other things, we now use the rootsolver
program called MPSolve \cite{mpsolve} to augment the internal rootsolvers in
Maple and Mathematica.

\begin{figure}
  \begin{center}
    \includegraphics[height=6.6cm]{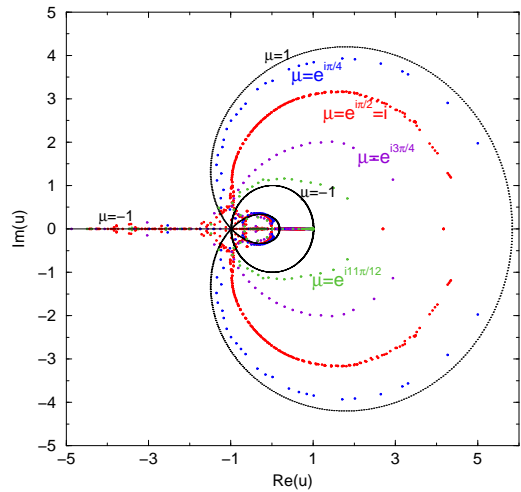}
  \end{center}
\caption{\footnotesize{Complex-$u$ zeros of the Ising model partition function
on sections of the square lattice for $\mu=e^{i\theta}$ with $\theta=\pi/4$,
$\pi/2$, $3\pi/4$, and $11\pi/12$.  See text for details of calculation. The
exact phase boundaries for $\theta=0$ and $\theta=\pi$ are also shown.}}
\label{sfbc_mu_circle}
\end{figure}

\begin{figure}
  \begin{center}
    \includegraphics[height=6.6cm]{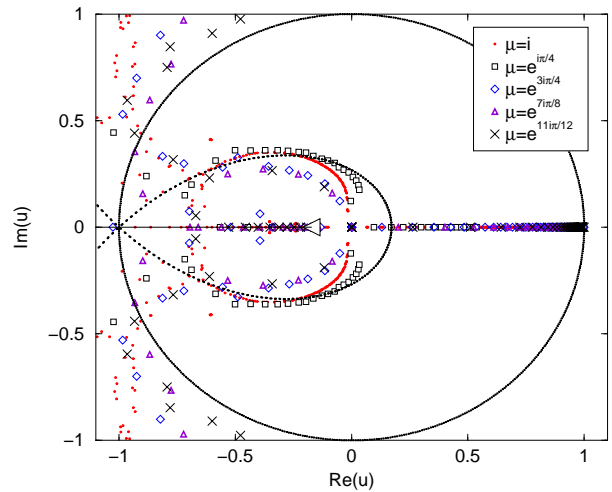}
  \end{center}
\caption{\footnotesize{A closer view of the complex-$u$ partition function
zeros in the region near $u=0$ for several values of $\theta$ in the range $0 <
\theta < \pi$. For reference the figure also shows, as exact results, the inner
part of the curve forming the lima\c{c}on (\ref{limacon}) for $\theta=0$ and
the unit circle on ${\cal B}_u$ for $\theta=\pi$. For visual clarity, for
$\theta=\pi$, we show only the right-hand endpoint, $u=-(3-2\sqrt{2})$, of the
real line segment on ${\cal B}_u$ (indicated with the symbol $\lhd$). The
lattice size is $12 \times 13$ except for $\theta=\pi/2$, for which we show
results with $L_x$ and $L_y$ up to 16.}}
\label{sfbc_mu_zoom}
\end{figure}

\begin{figure}
  \begin{center}
    \includegraphics[height=6.6cm]{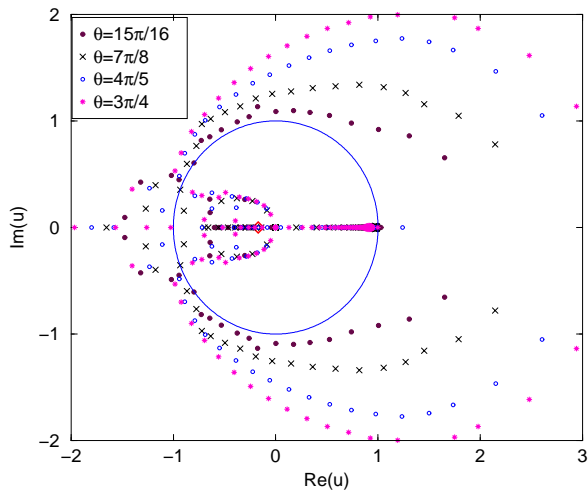}
  \end{center}
\caption{\footnotesize{A closer view of the complex-$u$ partition function
zeros in the inner central region for a several values of $\theta$ in the
range $\pi/2 < \theta < \pi$.  Lattice size is $12 \times 13$, as in
Fig. \ref{sfbc_mu_circle}.}}
\label{sfbc_theta_gt_half_zoom}
\end{figure}

\begin{figure}
  \begin{center}
    \includegraphics[height=6.6cm]{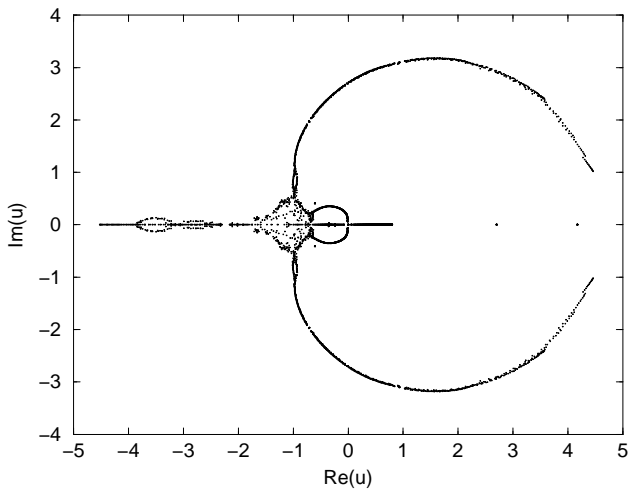}
  \end{center}
\caption{\footnotesize{Complex-$u$ zeros of the Ising model partition function
  with $\mu=e^{i\pi/2}=i$ on $L_x \times L_y$ sections of the square lattice
  with aspect ratio $\sim 1$ and $L_x$ and $L_y$ varying from 12 to 16.}}
\label{sfbc_muI}
\end{figure}

We show our results for the complex-$u$ zeros of the Ising model partition
function on $L_x \times L_y$ sections of the square lattice with
$\mu=e^{i\theta}$ in Fig. \ref{sfbc_mu_circle} for various values of $\theta$
in the range $0 < \theta < \pi$.  These zeros were calculated with the
symmetrized fixed boundary conditions defined above.  The curve (\ref{limacon})
for $\theta=0$ and the curve and line segment (\ref{bumum1}) for $\theta=\pi$
represent exact results. A more detailed view of the inner region near $u=0$ is
shown in Fig. \ref{sfbc_mu_zoom}.  We present a detailed view of the zeros in
the inner central region for $\pi/2 < \theta < \pi$ in
Fig. \ref{sfbc_theta_gt_half_zoom}.  The zeros presented for $\theta \ne \pi/2$
were calculated on $12 \times 13$ lattices.  As discussed further below, we
devoted a more intensive study to the value $\theta=\pi/2$, i.e., $\mu=i$, and
for this case we calculated the partition function and zeros for $L_x \times
L_y$ lattices with sizes $L_x$ and $L_y$ ranging from 12 to 16.  We show the
results for this $\mu=i$ case separately in Fig. \ref{sfbc_muI}.  Concerning
exact results, for visual clarity, in Figs. \ref{sfbc_mu_zoom} and
\ref{sfbc_theta_gt_half_zoom} for the case $\theta=\pi$, we show only the
right-hand endpoint of the real line segment (\ref{bumum1}) on ${\cal B}_u$ at
$u=-(3-2\sqrt{2})$ (indicated by the symbols $\lhd$ and $\Diamond$,
respectively).

As $\theta$ increases from zero, we observe a number of interesting features of
the complex-$u$ zeros of the partition function.  One important general feature
is that for $\theta \ne 0$ mod $\pi$, zeros occur on the real axis, extending
over an interval from the point where the inner loop of ${\cal B}_u$ is
inferred to cross this axis, to a right-hand endpoint $u_{rhe}$ that increases
as $\theta$ increases in the interval $0 < \theta < \pi$.  We infer that in the
thermodynamic limit (i) these zeros merge to form a real line segment on ${\cal
B}_u$ and (ii) the right-hand endpoint
\beq
u_{rhe} = e^{-4K_{rhe}} \equiv e^{-4J/(k_BT_{rhe})} 
\label{urhesq}
\eeq 
corresponds to the temperature $T_{rhe}$ at which the circular arc comprising
${\cal B}_\mu$ has endpoints at $\mu =e^{\pm i\theta}$.  For infinite
temperature, $K_{rhe}=0$, this endpoint (the Yang-Lee edge) occurs at
$\theta=\pi$ and as the temperature decreases, the endpoints of the circular
arc on ${\cal B}_\mu$ moves around to progressively smaller values of $\theta$.
As $T$ decreases to the critical temperature $T_c = T_{PM-FM}$ for the onset of
ferromagnetic long-range order, $\theta \to 0$, ${\cal B}_\mu$ closes to form
the unit circle $|\mu|=1$, and for lower temperatures it remains closed.  The
property that $u_{rhe}$ and the corresponding temperature $T_{rhe}$ increase
monotonically with $\theta$ in the range $0 < \theta < \pi$ is equivalent to
the property that the complex-conjugate endpoints of the circular arc
comprising ${\cal B}_\mu$ (i.e., the Yang-Lee edge) at $\theta$ increases
monotonically from $\theta=0$ at $T=T_c$ to $\theta \to \pi$ as $T \to
\infty$. This thus establishes a 1--1 correspondence between the right-hand
endpoint $u_{rhe}$ of the real line segment for a given $\theta$ and the
temperature at which this $\theta$ is the value of the endpoint of the circular
arc on ${\cal B}_\mu$.  The limit $\theta \to 0$ involves special behavior, in
that this real line segment shrinks to zero and disappears. The limit $\theta
\to \pi$ is also special; again, the line segment on the positive real $u$ axis
disappears in this limit and is replaced by a single zero at $u=1$ with
multiplicity $n/2$, where $n$ denotes the number of sites on the lattice.  This
zero at $u=1$ gives rise to the term $(1/2)\ln(u-1)$ in the reduced free energy
at $\mu=-1$ \cite{yl,mw67,ih}.

Turning on a finite (uniform) magnetic field, whether real or complex, does not
remove the PM-AFM phase transition that occurs for sufficiently large negative
$K=\beta J$.  It follows that the outer loop on ${\cal B}_u$ cannot break.  As
discussed in \cite{only}, this can be shown via a proof by contradiction.
Assume that this outer loop on ${\cal B}_u$ did break; then one could
analytically continue from the region around $K=0$, i.e., $u=1$, where the
staggered magnetization $M_{st}$ vanishes identically, to the physical AFM
phase where $M_{st}$ is nonzero, and similarly to the complex-$u$ extension of
this AFM phase, which would be a contradiction.  We find that as $\theta$
increases from 0 to $\pi$, the zeros that form the outer loop of ${\cal B}_u$
in the ${\rm Re}(u) > 0$ half-plane move monotonically inward toward the unit
circle $|u|=1$, which they form for $\theta \to \pi$.  We infer that in the
thermodynamic limit, (i) the right-most crossing on ${\cal B}_u$ decreases
monotonically from $u=u_{PM-AFM}=(3+2\sqrt{2}) \simeq 5.83$ to $u=1$ as
$\theta$ increases from 0 to $\pi$; and (ii) the upper and lower points where
the outer loop of ${\cal B}_u$ crosses the imaginary $u$ axis move
monotonically inward from $u=\pm (2 + \sqrt{3})i$ to $u=\pm i$.  From
inspection of the actual zeros that we calculate for various values of
$\theta$, we infer the following approximate maximal values of $u$ at which
${\cal B}_u$ crosses the positive real $u$ axis: $u \simeq 5.5$ for
$\theta=\pi/4$, $u \simeq 4.6$ for $\theta=\pi/2$, and $u \simeq 3.3$ for
$\theta=3\pi/4$.  

Recall that for nonzero real $h$, a theorem \cite{lp} guarantees that the free
energy for the ferromagnet is analytic for all temperatures, which means that
${\cal B}_u$ must break and retract from the real axis in the vicinity of what
was, for $h=0$ the PM-FM phase transition point, $u_{PM-FM}$ \cite{lp}.  In
contrast, the results of \cite{yl,lp} allow the free energy to be non-analytic
as $H$ is varied at constant $\beta$ or $\beta$ is varied at constant $H$
(i.e., in both cases, as $h$ is varied) if $Re(h)=0$, that is, $h$ is pure
imaginary, which is the situation that we consider here. Our most detailed
study of the partition function zeros, for $\mu=i$ (see Fig. \ref{sfbc_muI}) is
consistent with the conclusion that the inner loop on ${\cal B}_u$ does not
break but remains closed.  This conclusion is also consistent with our results
for other values of $\theta$.  In making this statement, we note that the fact
that the zeros on the right-hand side of the inner loop, calculated on finite
lattices, do not extend all the way in to the real axis, does not constitute
evidence of a break in this loop in the thermodynamic limit.  For example, even
for the exactly solved case $\mu=1$, the zeros calculated on finite lattices
also do not extend all of the way down to the real axis.  In this context, we
also remark on our exact results for quasi-1D strips; on both the toroidal and
cyclic ladder strips, for $\mu=e^{i\theta}$, as $\theta$ is increased from 0 to
$\pi$, the loop on ${\cal B}_u$ that passes through the critical point (at
$u=0$) remains intact and unbroken.  (Note that the point at which this loop
crosses the real axis for these quasi-1D infinite-length ladder strips remains
at $u=0$ as $\theta$ increases from 0 to $\pi$, while for the model in 2D the
inferred crossing point of the inner loop moves gradually to the left as
$\theta$ increases through this range for the square lattice.)

The details of the pattern of zeros in the complex-$u$ region that includes the
real interval $-1 \le u \le 0$ are complicated, and there is significant
scatter of some of these zeros.  Consequently, we do not try to make further
inferences about the form of the complex-$u$ boundary ${\cal B}_u$ in this
region in the thermodynamic limit.  As an example of the kind of feature that
might be present in this limit, one can discern some indication of possible
triple points at $u \simeq -0.7 \pm 0.2i$ and $u \simeq -0.9 \pm 0.5i$.  A
complex-conjugate pair of triple points is, indeed, present in our exact
solution for ${\cal B}_u$ on the toroidal ladder strip with $\mu=i$, as shown
in Fig. \ref{tormui}.  The scatter of zeros in this region raises the question
of whether some part of ${\cal B}_u$ might actually fill out 2-dimensional
areas rather than being one-dimensional (comprised of curves and possible line
segments) in the thermodynamic limit.  For the 2D Ising model in zero field it
is easy to see that complex-$u$ zeros generically fill out areas if the
spin-spin exchange constants in the $x$ and $y$ directions are unequal, but
this is not directly relevant here, since we only consider the model with
isotropic couplings.  For isotropic couplings, this area behavior happens for a
heteropolygonal Archimedean lattice, namely the $4 \cdot 8 \cdot 8$ lattice
\cite{cmo}, and here again, the origin of this is obvious from the exact form
of the free energy (see eq. (6.5) and Fig. 7 of \cite{cmo}).  One can fit
curves or line segments to many of the zeros in Fig. \ref{sfbc_mu_circle}.  As
for the region where the zeros show scatter, our results are not conclusive,
and we do not try to make any inference about whether or not some set of these
zeros might merge to form areas in the thermodynamic limit.

Several other aspects of the real zeros are of interest.  First, we find that
as $\theta$ increases from 0, there are real zeros not just to the right of the
extrapolated point where the inner loop on ${\cal B}_u$ crosses the real axis,
but also to the left of this point.  Indeed, we find that for $0 < \theta \le
\pi$, there are zeros on the negative real axis.  As $\theta \to \pi$, these 
occur in the interval of eq. (\ref{bumum1}), i.e., $-(3+2\sqrt{2}) \le u \le
-(3-2\sqrt{2})$. On the finite lattices that we have studied, we also have
found complex-conjugate pairs of zeros that are close to the zeros on the real
axis.

The value $\theta=\pi/2$, i.e., $\mu=i$, is the middle of the range under
consideration here, and we have devoted a particularly intensive study to it.
In addition to the general plots comparing the zeros for this value of $\theta$
with those for other values of $\theta$, we show the zeros for this case alone
in Fig. \ref{sfbc_muI}.  The zeros shown in this figure were calculated for
several different lattice sizes with aspect ratio $L_y/L_x \simeq 1$ and $L_x$
ranging from 12 to 16.  We display the zeros for different lattice sizes
together to see lattice size-dependent effects. As is evident from the figure,
much of the locus comprised by these zeros is largely independent of lattice
size for sizes this great.  These calculations illustrate our general
description of the properties of ${\cal B}_u$ above.  From these zeros we infer
that for $\mu=i$, in the thermodynamic limit, (i) the complex-$u$ phase
boundary ${\cal B}_u$ crosses the real axis at $u \simeq 4.6$ and the imaginary
axis at $u \simeq \pm 2.7i$; (ii) there is an inner loop on ${\cal B}_u$ which
is likely to pass through $u=0$, although there is some decrease in the density
of complex zeros in the vicinity of this point; (iii) the locus ${\cal B}_u$
exhibits a line segment on the real axis that extends from a right-hand
endpoint $u_{rhe} \simeq 0.8$ leftward with components along the negative real
axis; and although the details of this line segment at intermediate points
cannot be inferred with certitude, the left-most endpoint occurs at $u_{\ell
he} \simeq -4.5$; (v) the phase boundary ${\cal B}_u$ thus appears to separate
the $u$ plane into at least four regions: (a) the AFM phase and its complex-$u$
extension, which occupies values of $u$ going outward to complex infinity, (b)
the interior of the outer loop; (c) and the complex-conjugate pair of regions
inside of the inner loop, which seems to be divided into an upper and lower
part by the real line segment inside this loop.  As regards item (i), the
points at which the outer loop of ${\cal B}_u$ crosses the imaginary $u$ axis
are consistent, to within the accuracy of our calculations, with being equal to
$\pm (1+\sqrt{3})i$.

\subsection{Connections with Results on Quasi-1D Strips for $\mu=e^{i\theta}$}

With appropriate changes to take account of the change in dimensionality, we
can relate these features to our exact results on quasi-1D strips.  For these
strips we found that the locus ${\cal B}_u$ includes a line segment on the
positive real axis in the physical ferromagnetic interval as $\theta$ increases
above zero.  In the 1D case, we found the simple result $u_{rhe} =
\sin^2(\theta/2)$ in eq. (\ref{usegmentmucircle}) for the right-hand endpoint
of this line segment.  For the toroidal and cyclic ladder strips we
illustrated, e.g. for $\theta=\pi/2$ ($\mu=i$), how it is determined as
the root of the respective polynomials, eqs. (\ref{upolmui}) and
(\ref{cycrootpol}), that occur in the solution of a relevant cubic equation for
the eigenvalue of the transfer matrix.  In Table \ref{urhe_theta} we showed the
values of $u_{rhe}$ for the 1D line and the toroidal strip, together with the
corresponding values of the temperature $T_{rhe}$, as a function of $\theta
\equiv \theta_e$.  This table shows how $u_{rhe}$ and $T_{rhe}$ increase as
$\theta$ increases above 0 and approaches $\pi$.  We also noted how, for a
given value of $\theta$, $T_{rhe}$ increases as one increases the strip width.
This is physically understandable, since a given value of the angle $\theta$
corresponds to a higher temperature and hence larger $u_{rhe}$ as the strip
width increases because that increase fosters short-range ferromagnetic
ordering.

Another property of the zeros that can be related to our exact results on
quasi-1D strips is the part of the line segment extending to the left of the
point where the inner loop appears to cross the real axis and, indeed,
extending to negative real values.  For the quasi-1D strips, these intervals
are the same, since the ferromagnetic critical point is at $u=0$.  For the 1D
line case, the locus ${\cal B}_u$, which is $u \le \sin^2(\theta/2)$, includes
the semi-infinite line segment $u < 0$. For the toroidal ladder strip, we find
that for any $\theta$ in the interval $0 < \theta < \pi$, ${\cal B}_u$ includes
the segment $-1 \le u \le 0$ as well as the portion on the positive real axis
discussed above.  So there are again similarities with respect to this feature
as regards the results for the strips and for our zeros calculated on patches
of the square lattice.  In future work it would be of interest to calculate
complex-$u$ zeros of the Ising model partition function with imaginary $h$ on
$d=3$ lattices, as was done for real $h$ in Ref. \cite{ipz}.

\section{Connection of Singular Behavior of the Zero Density in
  the $\mu$ and $u$ Planes}

For the ferromagnetic Ising model, studies have been carried out of the
singularity at the endpoint of the circular arc ${\cal B}_\mu$ as $\theta \to
\theta_e$ (Yang-Lee edge) and the associated density of zeros $g(\theta)$ in
the original papers \cite{yl} and in works including those by Griffiths, Fisher
and collaborators, and Cardy \cite{kg}-\cite{cardy85}.  Kim has suggested that
the singularity at an edge of a locus of zeros in the $u$ plane is equivalent
to the Yang-Lee edge singularity \cite{kim3d}.  For the case of $d=2$
dimensions, we can show this equivalence using conformal field theory methods.
We recall that for a conformal field theory indexed by (relatively prime)
positive integers $p$ and $p'$, the central charge $c$ is
\beq
c = 1 - 6\frac{(p-p')^2}{p p'}
\label{cform}
\eeq
with scaling dimensions
\beq
h_{r,s} = \frac{[(pr-p's)^2-(p-p')^2]}{4pp'} \ , 
\label{hrs}
\eeq
where $1 \le r \le p-1$ and $1 \le s \le p'-1$.  The requirement of a single
scaling field (other than the identity) leads uniquely to the identification of
the conformal field theory as ${\cal M}_{5,2}$, which is non-unitary, with
central charge $c=-22/5$ \cite{cardy85}.  The scaling dimension for the single
non-identity field is $\eta = 4h_{1,2}=-4/5$.  From this and the standard
relation $\eta = d+2-2y_h$, where $y_h$ is the magnetic exponent, it follows
that $y_h = 12/5$.  Since the theory has only one relevant scaling field, the
thermal exponent is the same, i.e.
\beq
y_t = y_h = \frac{12}{5} \ . 
\label{ytyh}
\eeq
Substituting the value of $\eta$ into the scaling relation (\ref{sigma}) with
$d=2$ yields the result $\sigma=-1/6$ \cite{fisher78,cardy85}.  The equal
thermal and magnetic exponents in eq. (\ref{ytyh}) determine all of the rest of
the exponents for this critical point, which include
\beq
\nu_e' = \frac{1}{y_t}=\frac{5}{12}
\label{nue}
\eeq
\beq
\alpha_e' = 2 - \frac{d}{y_t} = \frac{7}{6}
\label{alphae}
\eeq
\beq
\beta_e = \frac{d-y_h}{y_t} = -\frac{1}{6}
\label{betae}
\eeq
\beq
\gamma_e' = \frac{2y_h-d}{y_t} = \frac{7}{6} \ . 
\label{gammae}
\eeq
(These results have also been independently and simultaneously obtained in this
manner by B. McCoy.)  The fact that the conformal field theory has only a
single non-identity operator and equal thermal and magnetic exponents leads to
the conclusion that the exponent $1-\alpha_e'$ describing the singular behavior
of the density of zeros at the right-hand endpoint $u_e=u_{rhe}$ of the
positive real line segment on ${\cal B}_u$ corresponding to a value of $\theta$
with $\mu=e^{i\theta}$, $0 < \theta < \pi$, is the same as the exponent
$\sigma$ describing the singular behavior of the density of zeros at the
complex-conjugate endpoints of the circular arcs on ${\cal B}_\mu$.

A parenthetical remark may be of interest here.  We recall that a liquid-gas
phase transition may be modelled as a lattice gas, and the latter may, in turn,
be mapped onto a ferromagnetic Ising model.  In this context the Yang-Lee
circle theorem states that the zeros of the partition function occur on the
unit circle $|z|=1$ in the complex plane of the fugacity, $z=e^{\beta \mu'}$
(where $\mu'$ is the chemical potential) and, in the thermodynamic limit, form
an arc of this unit circle extending from $\theta=\pi$ over to
complex-conjugate endpoints at $z=e^{\pm i\theta_e}$ \cite{yl}. For closed-form
approximations to the equation of state of a liquid-gas system such as that of
van der Waals, the density of Yang-Lee zeros has been calculated \cite{ky,hh}.
For intermolecular potentials with a repulsive hard core, the expansion of the
reduced pressure $pv_0/(k_BT)$ in powers of fugacity, $pv_0/(k_BT) =
\sum_{j=1}^\infty b_\ell z^\ell$, exhibits alternating signs, indicating a
singularity on the negative real $z$ axis \cite{groen}.  This latter
singularity has been shown to be equivalent to the Yang-Lee edge singularity
\cite{lf,pf}.  This is a somewhat different equivalence than the one discussed
here, since it relates the Yang-Lee singularities at $z=e^{\pm i\theta_e}$ to
one on the negative real $z$ axis, whereas the relation discussed here is
between the Yang-Lee singularities and a singularity on the (positive) real $u$
axis.

Now consider a switch from the imaginary values of $h$ relevant for the
Yang-Lee edge singularity to real values of $h$.  These lead to the
complex-conjugate arcs on ${\cal B}_u$ with arc (prong) endpoints $u_e$ and
$u_e^*$ that retract from the position of what was the critical point at
$u=u_c$ (for $h=0$) as $|h|$ increases from zero \cite{only}. Again, the fact
that the conformal field theory has only a single non-identity operator and
equal thermal and magnetic exponents leads us to the further conclusion that
the exponent $1-\alpha_e'$ describing the singular behavior of the zeros at the
ends of the complex-conjugate arcs in the complex $u$ plane at $u_e$ and
$u_e^*$ (cf. eq. (\ref{galpha}) with $u_s=u_e$) is the same as the exponent
$\sigma$.  With $\alpha_e'=7/6$, this implies that for the 2D Ising model,
$g(u)$ thus diverges at these arc endpoints with the exponent
\beq
1-\alpha_e' = \sigma = -\frac{1}{6}  \quad (2D) , 
\label{1minusalphae}
\eeq
We also conclude that the exact values of the exponents $\alpha_e'$, $\beta_e$,
and $\gamma_e'$ for the specific heat, magnetization, and susceptibility in
eqs.  (\ref{alphae})-(\ref{gammae}) apply to the arc endpoints at $u_e$ and
$u_e^*$.  These exact values agree very well with the numerical values that we
obtained in \cite{only} from our analysis of low-temperature, high-field (i.e.,
small-$u$, small-$\mu$) series (see Table I of \cite{only}).  These values had
been suggested in \cite{kim3d} based on the assumption that $y_t=y_h=12/5$ for
the endpoint of a locus of zeros in the $u$ plane.  Here we have proved the
equivalence using conformal field theory methods.

In \cite{only} we also studied complex values of $h$ corresponding to negative
$\mu$ in the real interval $-1 < \mu < 0$.  For the solvable case $\mu=-1$ one
knows the exponents $\alpha'_s$ and $\beta_s$ exactly at various singular
points, and in \cite{ih}, from analyses of series, we obtained the exponent
$\gamma'_s$ and inferred exact values for this exponent also.  These singular
points at $\mu=-1$ include the multiple point $u=-1$, the left- and right-hand
endpoints of the line segment $u_{\ell h e}=-(3+2\sqrt{2})$ and $u_{rhe} =
1/u_{\ell he}=-(3-2\sqrt{2})$, and also the point $u=1$. For reference, in
\cite{ih} we obtained $\alpha_e'=1$, $\beta_e=-1/8$, and $\gamma_e'=5/4$ at
$u_e=-(3-2\sqrt{2})$, $\alpha_s'=0$ ($C_H$ finite), $\beta_s=1/2$, and
$\gamma_s'=1$ at $u=-1$, and $\alpha_s'=0$ ($C_H$ finite), $\beta_s=-1/4$, and
$\gamma_s'=5/2$ at $u=1$ (see Table 4 of that paper), where the results for
$\alpha'$ and $\beta$ were exact and the results for $\gamma'$ were inferred
from our analysis of series. (Exponents at $u_{rhe} = 1/u_{\ell he}$ are
related by the $u \to 1/u$ symmetry.)  For $\mu$ close, but not equal, to
$\mu=-1$, the line segment on the negative real axis shifted slightly, and
there appeared a new line segment on the positive real axis extending inward
from the right-most portion of the boundary ${\cal B}_u$.  We also studied
these singular exponents via series analyses in \cite{only}.  In future work it
would be worthwhile to understand better the values of the exponents describing
these singularities for negative real $\mu$.

\section{Conclusions}

In this paper we have studied properties of the Ising model in
the complex $u$ plane for nonzero magnetic field.  We used exact results for
infinite-length quasi-1D strips to provide insights into the previous results
that we had obtained in \cite{ih,only}.  We also studied the case of complex
$h$, in particular, the case of imaginary $h$, for which $\mu=e^{i\theta}$.  We
used both exact results on strips and partition function zeros to analyze the
phase diagram in the $u$ plane for this range of $\mu$.  One important result
that we found was that the boundary ${\cal B}_u$ contains a real line segment
extending through part of the physical ferromagnetic interval $0 \le u \le 1$,
with a right-hand endpoint $u_{rhe}$ at the temperature for which the Yang-Lee
edge singularity occurs at $\mu=e^{\pm i\theta}$.  We also used conformal field
theory arguments to relate the singularities at $u_{rhe}$ and the Yang-Lee
edge.

\begin{acknowledgments}

We thank B. McCoy for a number of valuable and stimulating discussions and
I. Jensen and J.-M. Maillard for helpful comments.  The research of V.M. and
R.S. was partially supported by the grants NSF-DMS-04-17416 and
NSF-PHY-06-53342.

\end{acknowledgments}


\begin{thebibliography}{99}


\bibitem{ons}
L. Onsager, Phys. Rev. {\bf 65}, 117 (1944).

\bibitem{yang}
C. N. Yang, Phys. Rev. {\bf 85}, 808 (1952).

\bibitem{mwbook}
B. McCoy and T. T. Wu, {\it The Two-Dimensional Ising Model} (Harvard
University Press, Cambridge, 1968). 

\bibitem{yl}
C. N. Yang and T. D. Lee, Phys. Rev. {\bf 87}, 404 (1952);
T. D. Lee and C. N. Yang, Phys. Rev. {\bf 87}, 410 (1952).

\bibitem{fisher65}
M. E. Fisher, {\it Lectures in Theoretical Physics}
(Univ. of Colorado Press, Boulder, 1965), vol. 7C, p. 1.

\bibitem{dg}
C. Domb and A. J. Guttmann, J. Phys. C {\bf 3}, 1652 (1970). 

\bibitem{mw67}
B. M. McCoy and T. T. Wu, Phys. Rev. {\bf 155}, 438 (1967).

\bibitem{linwu}
K. Y. Lin and F. Y. Wu, Int. J. Mod. Phys. B {\bf 4}, 471 (1988). 

\bibitem{ih}
V. Matveev and R. Shrock, J. Phys. A {\bf 28}, 4859 (1995).

\bibitem{only}
V. Matveev and R. Shrock, Phys. Rev. E {\bf 53}, 254-267 (1996). 

\bibitem{yy}
V. Matveev and R. Shrock, Phys. Lett. A {\bf 215}, 271 (1996). 

\bibitem{tlow}
M. F. Sykes, D. S. Gaunt, J. L. Martin, S. R. Mattingly, and
J. W. Essam, J. Math. Phys. {\bf 14}, 1071 (1973); 
M. F. Sykes, M. G. Watts, and D. S. Gaunt, J. Phys. A {\bf 8}, 1448 (1975).

\bibitem{be}
R. J. Baxter and I. G. Enting, J. Stat. Phys. {\bf 21}, 103 (1979).

\bibitem{kim}
S.-Y. Kim, Phys. Rev. E {\bf 71}, 017102 (2005). 

\bibitem{newp}
I. Jensen, J.-M. Maillard, V. Matveev, B. M. McCoy, and R. Shrock, work in
progress, to appear.

\bibitem{a}
R. Shrock, Physica A {\bf 283}, 388 (2000).

\bibitem{ta}
S.-C. Chang and R. Shrock, Physica {\bf A 286}, 189 (2000).

\bibitem{hca}
S.-C. Chang and R. Shrock, Physica A {\bf 296}, 183 (2001).

\bibitem{abe}
R. Abe, Prog. Theor. Phys. {\bf 37}, 1070 (1967); 
Prog. Theor. Phys. {\bf 38}, 322 (1967). 

\bibitem{chisq}
V. Matveev and R. Shrock, J. Phys. A {\bf 28}, 1557 (1995).

\bibitem{blomberg1}
O. Stormark and C. Blomberg, Physica Scripta {\bf 1}, 47 (1970).

\bibitem{blomberg2}
C. Blomberg, Physica Scripta {\bf 2}, 117 (1970).

\bibitem{suz67}
M. Suzuki, Prog. Theor. Phys. {\bf 38}, 1225 (1967). 

\bibitem{glumac} 
Z. Glumac and K. Uzelac, J. Phys. A {\bf 27}, 7709 (1994). 

\bibitem{lp}
J. L. Lebowitz and O. Penrose, Commun. Math. Phys. {\bf 11}, 99 (1968).

\bibitem{bk}
H. J. Brascamp and H. Kunz, J. Math. Phys. {\bf 15}, 65 (1974).

\bibitem{mpsolve}
D. A. Bini and G. Fiorentino, Numerical Algorithms {\bf 23}, 127 (2000);
MPSolve program available at \goodbreak 
http://www.dm.unipi.it/cluster-pages/mpsolve/index.htm . 

\bibitem{cmo}
V. Matveev and R. Shrock, J. Phys. A {\bf 28}, 5235 (1995).  

\bibitem{ipz}
C. Itzykson, R. B. Pearson and J.-B. Zuber, Nucl. Phys. B {\bf 220}, 415
 (1983).

\bibitem{kg}
P.J. Kortman and R.B. Griffiths, Phys. Rev. Lett. {\bf 27}, 1439 (1971).

\bibitem{kf}
D.A. Kurtze and M.E. Fisher, Phys. Rev. B {\bf 20}, 2785 (1979).

\bibitem{fisher78}
M. E. Fisher, Phys. Rev. Lett. {\bf 40}, 1610 (1978).
                                             
\bibitem{cardy85}
J. L. Cardy, Phys. Rev. Lett.. {\bf 54}, 1354-1356 (1985).

\bibitem{kim3d}
S.-Y. Kim, Nucl. Phys. B {\bf 637}, 409 (2002). 

\bibitem{ky}
S. Katsura, J. Chem. Phys. {\bf 22}, 1277 (1954). 

\bibitem{groen}
J. Groeneveld, Phys. Lett. {\bf 3}, 50 (1962). 

\bibitem{hh}
P. C. Hemmer and E. Hiis Hauge, Phys. Rev. {\bf 133}, A1010 (1964).

\bibitem{lf}
S.-N. Lai and M. E. Fisher, J. Chem. Phys. {\bf 103}, 8144 (1995). 

\bibitem{pf}
Y. Park and M. E. Fisher, Phys. Rev. E {\bf 60}, 6323 (1999). 

\end{thebibliography}
\end{document}